\documentclass[journal,twoside,web]{IEEEtran}

\usepackage{tmi}
\usepackage{cite}
\usepackage{amsmath,amssymb,amsfonts}
\usepackage{algorithmic}
\usepackage{graphicx}
\usepackage{textcomp}
\usepackage{graphicx}
\usepackage{xcolor}
\usepackage{multirow}
\usepackage{tabu}
\usepackage{svg}
\usepackage{bbm}
\usepackage{comment}
\usepackage{array}
\usepackage{amstext}
\usepackage{makecell}
\usepackage{caption}
\usepackage{tablefootnote}
\usepackage{booktabs} 
\usepackage{float}

\usepackage{colortbl}

\usepackage{placeins}
\usepackage{nicefrac} 

\newcommand{\ie}{\textit{i.e.}}

\newcommand{\etal}{\textit{et al.}}

\usepackage[pagebackref=false,breaklinks=true,colorlinks,bookmarks=false]{hyperref}

\def\BibTeX{{\rm B\kern-.05em{\sc i\kern-.025em b}\kern-.08em
    T\kern-.1667em\lower.7ex\hbox{E}\kern-.125emX}}
\begin{document}
\title{Deep Rib Fracture Instance Segmentation and Classification from CT on the RibFrac Challenge}
\author{Jiancheng Yang, Rui Shi, Liang Jin, Xiaoyang Huang, Kaiming Kuang, Donglai Wei, Shixuan Gu, Jianying Liu, Pengfei Liu, Zhizhong Chai, Yongjie Xiao, Hao Chen, Liming Xu, Bang Du, Xiangyi Yan, Hao Tang, Adam Alessio, Gregory Holste, Jiapeng Zhang, Xiaoming Wang, Jianye He, Lixuan Che, Hanspeter Pfister, Ming Li, Bingbing Ni
\thanks{This work was supported by National Science Foundation of China (U20B200011, 61976137). This work was also supported by Grant YG2021ZD18 from Shanghai Jiao Tong University Medical Engineering Cross Research. Jiancheng Yang and Rui Shi contributed equally. Corresponding author: Bingbing Ni (e-mail: nibingbing@sjtu.edu.cn).}
\thanks{J. Yang, R. Shi, X. Huang, B. Ni are with Shanghai Jiao Tong University, Shanghai, China. }
\thanks{L. Jin is with Radiology Department, Huadong Hospital affiliated to Fudan University, Shanghai, China and with Huashan Hospital affiliated to Fudan University, Shanghai, China and also with Shanghai Key Lab of Forensic Medicine, Key Lab of Forensic Science, Academy of Forensic Science, Ministry of Justice, China.}
\thanks{K. Kuang is with Dianei Technology, Shanghai, China, and also with UC San Diego, CA, USA. }
\thanks{D. Wei is with Boston College, MA, USA. }
\thanks{S. Gu is with Harvard University, MA, USA.} 
\thanks{J. Liu, P. Liu are with Huiying Medical Technology (Beijing) Co., Ltd., Beijing, China. }
\thanks{Z. Chai, Y. Xiao are with ImsightMed, Shenzhen, China.}
\thanks{H. Chen is with Hong Kong University of Science and Technology, Hong Kong SAR. }
\thanks{L. Xu and B. Du are with Zhejiang University, Hangzhou, China. }
\thanks{X. Yan, H. Tang are with UC Irvine, CA, USA.}
\thanks{A. Alessio is with Michigan State University, MI, USA. }
\thanks{Gregory Holste is with University of Texas at Austin, TX, USA, and also with Michigan State University, MI, USA. }
\thanks{J. Zhang, X.Wang are with University of Shanghai for Science and Technology, Shanghai, China.}
\thanks{J. He, L. Che are with DeepBlue Technology (Shanghai) Co., Ltd, Shanghai, China. }
\thanks{H. Pfister is with Harvard University, MA, USA.}
\thanks{M. Li is with Radiology Department, Huadong Hospital affiliated to Fudan University, Shanghai, China, and also with Institute of Functional and Molecular Medical Imaging, Shanghai, China.}
}

\maketitle


\begin{abstract}

Rib fractures are a common and potentially severe injury that can be challenging and labor-intensive to detect in CT scans. AI has the potential to assist in identifying and diagnosing rib fractures, but the unique shape of each rib, with a diagonal course across numerous CT sections, presents a technical hurdle. While there have been efforts to address this field, the lack of large-scale annotated datasets and evaluation benchmarks has hindered the development and validation of deep learning algorithms.
To address this issue, the RibFrac Challenge was introduced, providing a benchmark dataset of over 5,000 rib fractures from 660 CT scans, with voxel-level instance mask annotations and diagnosis labels for four clinical categories (buckle, nondisplaced, displaced, or segmental). The challenge includes two tracks: a detection (instance segmentation) track evaluated by an FROC-style metric and a classification track evaluated by an F1-style metric. During the MICCAI 2020 challenge period, 243 results were evaluated, and seven teams were invited to participate in the challenge summary. The analysis revealed that several top rib fracture detection solutions achieved performance comparable or even better than human experts. Nevertheless, the current rib fracture classification solutions are hardly clinically applicable, which can be an interesting area in the future.
As an active benchmark and research resource, the data and online evaluation of the RibFrac Challenge are available at the challenge website (\url{https://ribfrac.grand-challenge.org/}). As an independent contribution, we have also extended our previous internal baseline by incorporating recent advancements in large-scale pretrained networks and point-based rib segmentation techniques. The resulting FracNet+ demonstrates competitive performance in rib fracture detection, which lays a foundation for further research and development in AI-assisted rib fracture detection and diagnosis.

\end{abstract}

\begin{IEEEkeywords}
rib fracture, 3D detection, 3D classification, 3D instance segmentation, computed tomography.
\end{IEEEkeywords}

\section{Introduction}
\label{sec:introduction}

\IEEEPARstart{R}{ib} fractures are a common injury that can result from various causes such as falls, trauma, athletic activities, non-accidental injury, or primary bone tumors and metastatic lesions~\cite{baiu2019rib,peek2020epidemiology,kuo2021rib}. Internal injuries such as liver or spleen lacerations, mediastinal injury, pneumothorax, hemothorax, flail chest, and pulmonary contusions may also be associated with rib fractures~\cite{ingoe2020epidemiology}. The severity of trauma can be indicated by the number of fractured ribs, which can result in increased morbidity and mortality rates~\cite{talbot2017traumatic,peek2020traumatic}. Counting the number of rib fractures is important in forensic examinations for determining the degree of disability~\cite{kolopp2020automatic,jin2018low}. 


Multidetector computed tomography (CT) scanning is a valuable tool for identifying rib fractures with higher accuracy than standard chest radiographs~\cite{banaste2018whole,jin2018low,peek2020traumatic}. However, the sheer volume of images generated by CT scans, coupled with the complex shape of each rib and its diagonal course across numerous CT sections, makes interpreting them challenging~\cite{ringl2015ribs}. This difficulty is compounded in cases of polytraumatized patients, where radiologists are under pressure to rapidly identify life-threatening injuries~\cite{banaste2018whole,blum2021automatic}. As a result, secondary injuries like rib fractures can be overlooked, with studies showing that up to 20.7\% of rib fractures may be missed on CT scans~\cite{pinto2016errors,blum2021automatic}. While most of these missed fractures may be minor, their consequences can still be significant for patients, clinicians, and radiologists. 
Among different types of fractures, buckle fractures are the most frequently missed type of fracture due to their confusing appearance, and nondisplaced fractures can also be missed when they are parallel to the scan plane of the CT images~\cite{dankerl2017evaluation,cho2012missed}. Diagnosing subtle fractures is a tedious and time-consuming process, requiring the sequential evaluation of a large number of CT slices, rib-by-rib and side-by-side~\cite{ringl2015ribs}. In cases where missed injuries are a concern, some experts have recommended double-reading whole-body CT scans in high-risk patients~\cite{banaste2018whole}. However, this may not always be feasible.

Emerging artificial intelligence (AI) technology has made significant strides in the field of medical services and delivery, particularly in screening~\cite{setio2017validation,mckinney2020international,yang2020alignshift,yang2021asymmetric,xu2022lssanet}, diagnosis~\cite{keane2018eye,zhao20183d,yang2019probabilistic,zhang2020clinically,yang2020hierarchical,ding2022improving}, and treatment~\cite{bi2019artificial,yang2020mia,yang2021multi,thies2022combined,deng2022genopathomic}. With high-performance deep learning, computer-aided diagnosis powered by AI can help reduce the burden on human labor, enhance diagnosis consistency and accuracy, personalize patient treatment, and improve the patient-doctor relationship~\cite{topol2019high}. 
AI can transform rib fracture diagnosis and treatment by aiding healthcare professionals. Automated diagnosis through AI, such as rib unfolding, streamlines the process and improves accuracy, though training is needed to avoid errors~\cite{ringl2015ribs,urbaneja2019automatic,kolopp2020automatic,blum20203d}. AI-assisted radiologists, particularly using convolutional neural networks, have shown improved accuracy in rib fracture detection~\cite{weikert2020assessment,zhou2020automatic,Jin2020DeeplearningassistedDA}. However, the lack of large-scale annotated datasets and evaluation benchmarks limits deep learning development in this area, underscoring the need for comprehensive datasets and benchmarks for AI model training, evaluation, and research advancement.



\begin{figure}
\centering
\includegraphics[width=.95\linewidth]{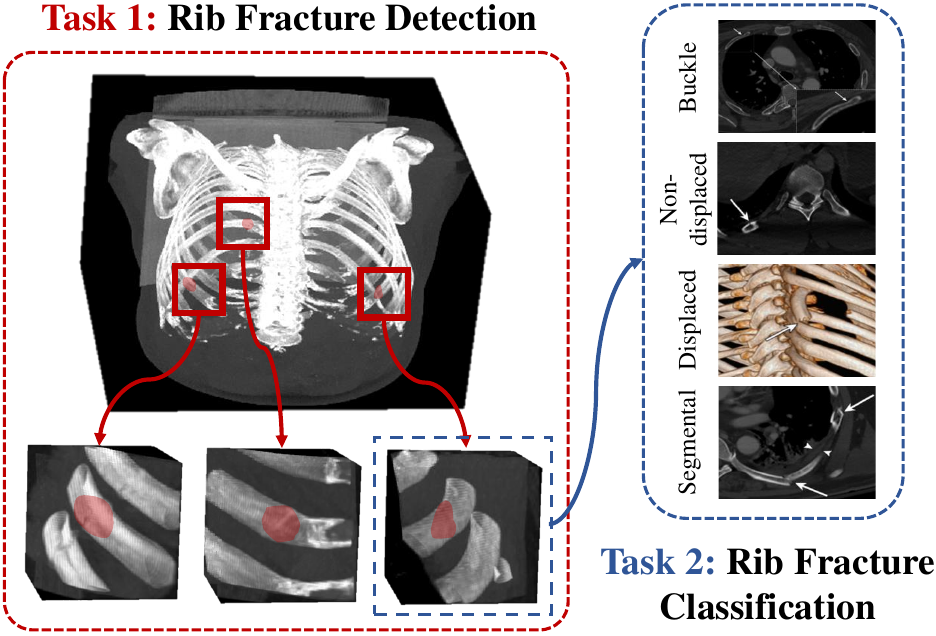}
\caption{\textbf{Illustration of two tracks in the RibFrac Challenge.} In the rib fracture detection track, participants submit the 3D instance segmentation mask for each fracture to detect their location and extent. In the rib fracture classification track, participants submit the label for each fracture to identify their type (buckle, non-displaced, displaced, or segmental). The red regions are the segmentation masks of rib fracture instances.}
\label{fig:track_illustration}
\end{figure}

The RibFrac Challenge was launched to facilitate the development and evaluation of automated methods for rib fracture detection and diagnosis from CT scans. The challenge provides the first large-scale benchmark dataset of rib fractures from CT scans, comprising over 5,000 rib fractures from 660 CT scans. Each scan has voxel-level instance mask annotations and diagnosis labels for four clinical categories: buckle, nondisplaced, displaced, or segmental. The challenge consists of two tracks: a detection track and a classification track, as illustrated in Fig.~\ref{fig:track_illustration}. In the detection (instance segmentation) track, participants are required to submit the 3D instance segmentation mask for each fracture to detect their location and extent. The detection performance is ranked using a free-response receiver operating characteristic (FROC)-based metric. Segmentation metrics (IoU and Dice) are also provided for analysis, but they are not used for ranking because the rib fracture detection is more important than segmentation in clinical practice. In the classification track, participants are required to submit the label for each fracture to identify their type based on their detection results. The classification performance of the end-to-end system (from detection to classification) is ranked using an F1 score-based metric. The dataset is the largest publicly available dataset of rib fractures with instance-level annotations, and offers a unique opportunity for researchers to develop and validate deep learning algorithms for rib fracture detection and diagnosis.

The challenge was held as a satellite event during MICCAI 2020, and a total of 243 submissions were evaluated during the official challenge period. From these, seven teams were selected to participate in the challenge summary. The analysis has demonstrated the potential of AI-assisted rib fracture detection and diagnosis, as some of the top detection solutions achieved performance comparable to or even better than human experts. However, the current rib fracture classification solutions are not yet clinically applicable, emphasizing the need for further research and development in this domain. The challenge, both historically and presently, is hosted on the \textit{Grand Challenge} website\footnote{\url{https://ribfrac.grand-challenge.org/}}, offering a comprehensive dataset and an online evaluation platform for advancing research and development in AI-assisted rib fracture detection and diagnosis. As of December 2023, it has attracted over 1,200 registered users. Post the official competition period, additional teams have submitted their results and elevated the benchmark. This indicates that the RibFrac Challenge is actively contributing to the development of deep learning in rib fracture diagnosis.

As an independent technical contribution distinct from the benchmark and challenge, this paper delineates how we have enhanced our internal baseline FracNet~\cite{Jin2020DeeplearningassistedDA}. Post the official competition, the field has witnessed significant advancements. For instance, we have developed a point-based rib segmentation technique on the RibFrac dataset~\cite{Yang2021RibSegDA,Jin2022RibSegVA}, which is expected to significantly improve rib fracture detection. 
Furthermore, the emergence and accessibility of large-scale pre-trained networks for 3D medical imaging have been notable~\cite{huang2023stu}. In light of these developments, we introduced FracNet+, which has achieved competitive results in rib fracture detection. This underscores the beneficial role of rib segmentation in the fracture detection and establishes a foundation for future research in this area.

Our contributions can be summarized as follows: 

\begin{itemize}
    \item \textbf{Formalizing research problem.} The RibFrac Challenge represents the first instance of formalizing rib fracture diagnosis as a machine learning problem. The task is well-designed, enabling participants to develop and assess automated methods specifically focused on rib fracture detection and diagnosis.
    \item \textbf{Large in scale.} The dataset stands as the largest publicly available collection of rib fractures with instance-level mask annotations. It comprises over 5,000 rib fractures from 660 CT scans, providing a rich dataset for researchers to train and validate deep learning algorithms.
    \item \textbf{Community impact.} With more than 1,200 registered users, the challenge has garnered considerable interest and participation from the research community. It has been instrumental in the evolution of deep learning algorithms for rib fracture detection and diagnosis.
    \item \textbf{Strong internal method.} Capitalizing on recent advancements in large-scale pretrained medical models and domain-specific progress in rib segmentation, we have extended the previous internal method for rib fracture detection and achieved competitive performance.
\end{itemize}

\section{Related Work}

\subsection{Rib Fracture Detection}

Several recent studies have explored deep learning approaches for rib fracture detection. For instance, a UNet model incorporating a dual-attention mechanism demonstrated improved performance compared to earlier methods. However, it had limited applicability as it only worked on a restricted number of CT slides and lacked a detailed annotation protocol for their dataset \cite{Zhou2022RibFD}. Classic models like DenseNet \cite{Huang2017DenselyCC} and ResNet \cite{7780459} have also been utilized for automatic rib fracture detection, but the models were not released, and they were evaluated on relatively small datasets that were not publicly available, which may affect the generalizability of their results \cite{yao2021rib,weikert2020assessment}. Some architectures typically used in object detection and image classification, such as CenterNet \cite{Duan2019CenterNetKT} and DLA34 \cite{Yu2017DeepLA}, have been applied to screening and false-positive elimination in rib fracture detection, but they did not include a large validation set~\cite{Yang2022DevelopmentAA}. Other studies have focused on the clinical validation of deep learning methods for rib fracture detection. For example, comparative studies have demonstrated that deep learning-based software can be integrated into radiology workflows to enhance rib fracture detection accuracy and reading efficiency \cite{Zhang2020ImprovingRF,liu2022clinical,Meng2021AFA}. Additionally, Kaiume~\etal{} utilized a deep convolutional neural network-based software and compared its rib fracture diagnostic performance with doctors \cite{Kaiume2021RibFD}, while Niiya~\etal{} evaluated the clinical effectiveness of deep learning models for rib fractures in high-energy trauma patients \cite{niiya2022development}. However, these studies were limited by their relatively small in-house datasets. While these works demonstrate the potential of deep learning-based approaches for rib fracture detection, they also highlight the need for larger, more diverse datasets with detailed annotations, and more rigorous validation to improve accuracy and clinical applicability.

\subsection{Rib Segmentation and Centerline Extraction}

Rib segmentation and anatomical centerline extraction play crucial roles in various clinical applications. Rib segmentation provides a stable reference for lung volume estimation \cite{Mansoor2014AGA} and bone abnormality quantification \cite{Fokin2018QuantificationOR}. Anatomical centerlines derived from rib structures enable the localization of organs for surgery planning \cite{Wang2005ART} and the registration of pathologies like lung nodules \cite{Shen2003ATC}. Furthermore, automatic rib segmentation and centerline extraction are essential for developing visualization tools for unfolded rib cages \cite{ringl2015ribs,Bier2015EnhancedRT, Abe2014HighspeedPC}. 

In recent years, several studies have investigated rib segmentation and centerline extraction methods using CT scans. The RibSeg dataset \cite{Yang2021RibSegDA,Jin2022RibSegVA} was specifically created to benchmark rib labeling and anatomical centerline extraction using CT scans from the RibFrac challenge. MDUNet \cite{Wang2020MDUNetAC} employs multiscale feature fusion and dense connections \cite{Huang2017DenselyCC} to segment clavicles and ribs. Other approaches include a detection-then-segmentation pipeline with nine degrees-of-freedom pose estimation \cite{Guo2022MedQuerySP} and trainable segmentation networks that combine multiple 3D UNets at different resolutions \cite{Schnider2023ImprovedDB} for rib segmentation from CT scans. Additionally, studies have focused on rib centerline extraction, employing techniques such as deformable template matching \cite{wu2012learning} and rib tracing \cite{lenga2018deep} on rib cages detected by deep learning models.

The significance of rib segmentation and centerline extraction could serve as structural references for rib fracture detection. Our internal method has successfully demonstrated its beneficial role.

\section{The RibFrac Challenge}

\subsection{Dataset}
\subsubsection{Data Acquisition and Pretreatment}


\begin{figure}
\centering
\includegraphics[width=.95\linewidth]{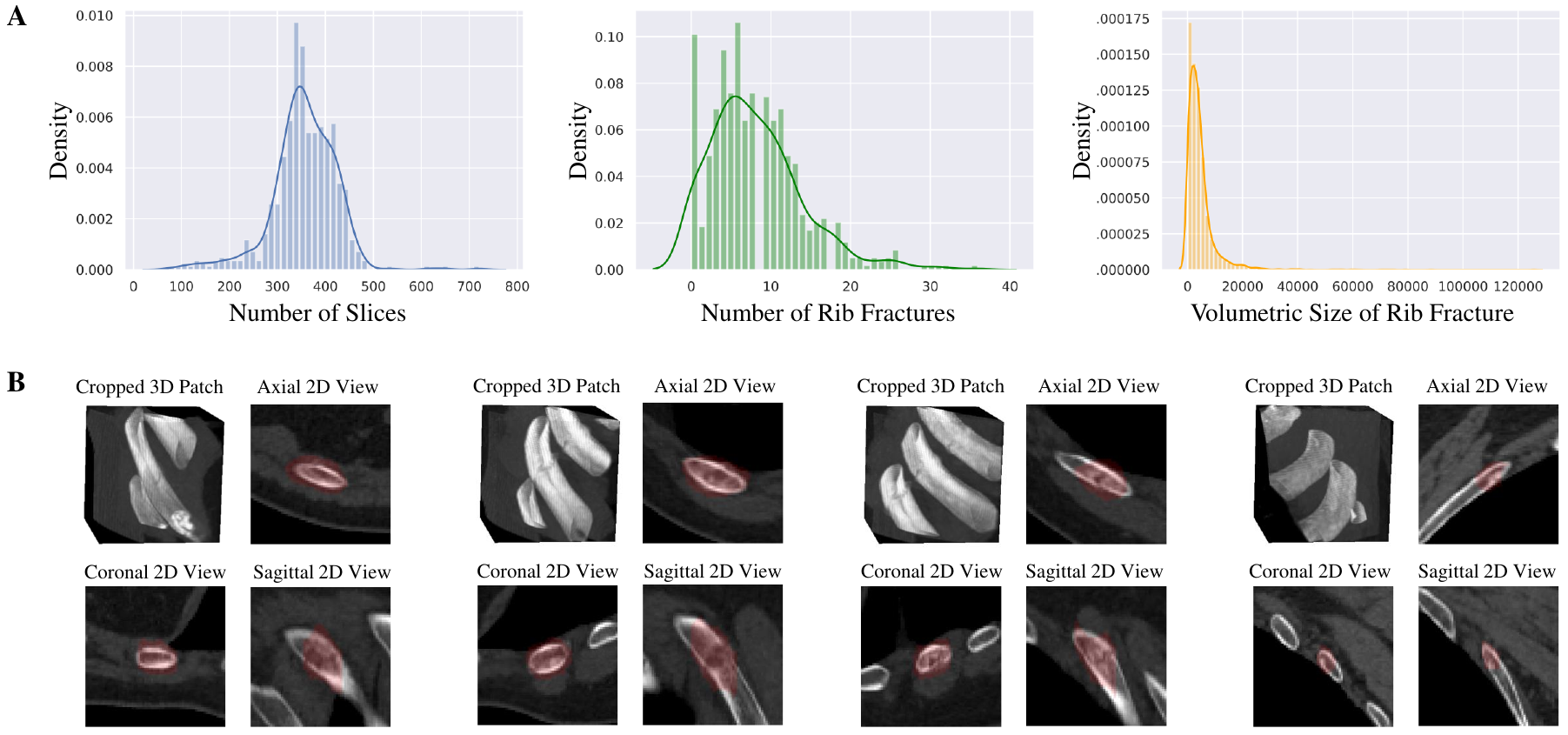}
\caption{\textbf{Statistics and sample visualization of the RibFrac Challenge dataset.} A) Histograms that display the distribution of the number of slices per scan (left), the number of rib fractures per scan (middle), and the volumetric size of each individual rib fracture (right). B) Visualization of four rib fracture samples. Each of the four rib fracture samples is represented by a cropped 3D CT patch using volumetric rendering (upper left). The axial (upper right), coronal (lower left) and sagittal (lower right) from the fracture centroid are shown in 2D, along with the human-annotated rib fracture voxel-level segmentation.}
\label{fig:ribfrac_dataset_illustration}
\end{figure}

\begin{table}
    \centering
    \caption{\textbf{RibFrac dataset subsets and fracture categories}. The table shows the number of patients and rib fractures in each subset of the RibFrac dataset. The RibFrac contains four different categories of rib fractures, including buckle (BK), non-displaced (ND), displaced (DP), and segmental (SG). Fractures that are difficult to categorize are labeled as unclassified (UN) and are excluded from the evaluation.}
    \label{tab:ribfrac_dataset_split}
    \begin{tabular}{l|cc|ccccc}
        \toprule
        Subset & Patients & Fractures & BK & ND & DP & SG & UN\\
        \midrule
        Training & 420 & 3,987 & 618 & 567 & 291 & 179 & 2,332  \\
        Validation & 80  & 435 & 69 & 63 & 30 & 30 & 243 \\
        Test & 160  & 882 & 151 & 188 & 84 & 50 & 409 \\
        \midrule
        Total & 660 & 5,304 & 838 & 818 & 405 & 259 & 2,984 \\
        \bottomrule
    \end{tabular}
\end{table}

The RibFrac Dataset used in the challenge has been specifically curated for rib fracture detection, segmentation, and classification. This comprehensive dataset consists of chest-abdomen CT scans, rib fracture segmentation, and classification information for a total of 660 patients. The data collection was conducted retrospectively and received approval from the ethics committee of Huadong Hospital affiliated to Fudan University (NO.2019K146), with informed consent waived.

To acquire the data, we utilized two advanced CT scanners: the 16cm wide coverage detector CT (Revolution CT, GE Healthcare, WI, USA) and the second-generation dual-source CT scanner (Somatom Definition Flash, Siemens Healthcare, Forchheim, Germany). The inclusion of data from two distinct CT scanners enhances the robustness and adaptability of the developed algorithms. To ensure patient privacy, the CT scans and rib fracture annotations were converted from the raw DICOM (Digital Imaging and Communications in Medicine) to the NIFTI (Neuroimaging Informatics Technology Initiative) format. We partitioned the dataset into three subsets: the training set (420 samples), the validation set (80 samples), and the test set (160 samples), where the labels for the test set were hidden for evaluation. For a detailed breakdown of the number of patients, CT slices, and rib fractures in each subset, please refer to Tab.~\ref{tab:ribfrac_dataset_split}. Additionally, you can find a visual representation of the RibFrac Dataset in Fig.~\ref{fig:ribfrac_dataset_illustration}, showcasing dataset statistics and sample visualizations.

\subsubsection{Data Labeling}

To ensure a fair evaluation in the RibFrac Challenge, five radiologists, labeled as A (3-5 years of experience), B (10-20 years), C (5 years), D (5 years), and E (20 years), participated in the labeling process. Initially, radiologists A and B examined the CT images within 48 hours of the examinations. Subsequently, radiologists C and D manually annotated the volume of interest for rib fractures based on the diagnosis reports compiled by A and B. The voxel annotations were performed using the 3D Slicer software (version 4.8.1, Brigham and Women's Hospital). The annotations from C and D were later verified by senior radiologist E.

To enhance the comprehensiveness of the annotations, we developed a sophisticated human-in-the-loop procedure that combined the assistance of a deep learning model and human expertise. A deep learning model, following the approach in~\cite{Jin2020DeeplearningassistedDA}, was trained using the initial training subset of the RibFrac dataset. This model predicted potential fracture region candidates. Predictions that were not covered by the initial labels were verified by radiologist E and added to the annotations if confirmed. Our estimation suggests that approximately 20\% rib fractures were missed during the initial labeling. It is important to note that no data leakage issue occurred in the human-in-the-loop labeling process.

In addition to the rib fracture segmentation masks, we also provide classification labels. Rib fractures are classified into four different classes (refer to Fig.~\ref{fig:track_illustration} for visual examples):

\begin{itemize}
\item \textbf{Buckle fractures (BK)} are incomplete fractures that appear as bulges in the rib. While buckle fractures are common in pediatric patients across various bones, they can be easily missed in radiology examinations.

\item \textbf{Non-displaced fractures (ND)} are fractures that do not cause any displacement or misalignment of the bones. Therefore, they can be challenging to identify in radiography. Non-displaced rib fractures may only be observed in follow-up exams when signs of healing have already manifested. Radiologists should be vigilant for associated injuries when there are no direct signs of such fractures in radiography.

\item \textbf{Displaced fractures (DP)} are those when there are cortical disruptions and significant abnormalities in alignment. Injuries to the surrounding tissues and structures may occur, and several lethal complications related to displaced rib fractures have been documented.

\item \textbf{Segmental fractures (SG)} are severe injuries characterized by at least two separate complete fractures in the same rib. Those affecting three or more contiguous rib levels are associated with an increased risk of flail chest.
\end{itemize}

Please note that we also have a considerable number of fractures classified as unknown (UN), indicating that they are rib fractures but their specific type could not be determined due to ambiguity or diagnostic difficulty. These fractures are excluded during the evaluation process. Tab.~\ref{tab:ribfrac_dataset_split} presents the statistics of rib fracture classification labels for the RibFrac training, validation and test subsets.


\begin{figure}
\centering
\includegraphics[width=\linewidth]{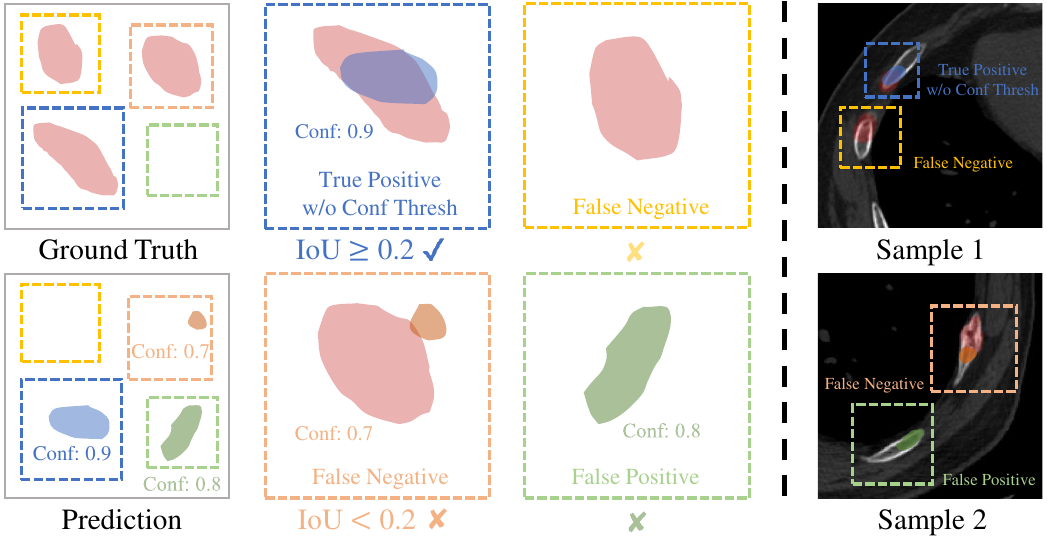}
\caption{\textbf{Illustration of detection hit in the FROC.} \textbf{Left:} A detection proposal is considered a hit (or a true positive without the confidence threshold) when it overlaps with any annotation with an IoU $\geq$ 0.2 (depicted in blue). The orange and yellow instances represent false negatives, indicating annotated fractures that were not detected. The green one denotes a false positive, indicating a predicted fracture that does not exist. \textbf{Right:} Two illustrative samples on CT slices.}
\label{fig:froc_illustration}
\end{figure}

\subsection{Evaluation}
\label{sec:evaluation}

There are two tracks in the RibFrac Challenge: the detection (instance segmentation) track and the classification track. In the detection track, participants are required to submit the 3D instance segmentation mask for each fracture, aiming to identify the location and extent of the fractures. In the classification track, participants need to submit the label for each fracture to determine its type based on the detection results. The labels for the training and validation sets are provided to the participants, allowing them to assess the model performance. However, the labels for the test set are withheld, and model performance can only be evaluated by submitting the results to the official website of the Grand Challenge. To ensure transparency and provide participants with detailed evaluation information, the evaluation code is made available\footnote{\url{https://github.com/M3DV/RibFrac-Challenge}}.

\subsubsection{Detection}

In this task, participants are given the task of detecting rib fractures from CT scans, and the evaluation is based on Free-Response Receiver Operating Characteristic (FROC) analysis for detection. Essentially, participants are engaged in 3D instance segmentation; however, for a more relevant assessment of the algorithms in terms of their practical applicability in clinical settings, we evaluate their performance using an FROC-style detection metric.

Due to the elongated shape of rib fractures, the detection task needs to be approached using instance segmentation. Each rib fracture instance is annotated by radiologists with a voxel-level mask that represents the fracture region. However, due to the inherent ambiguity of the fracture region, the instance masks may contain noise. Therefore, segmentation predictions are primarily used to calculate overlap in the detection metric.

The evaluation of the detection performance utilizes FROC analysis, which balances sensitivity and false positives (FP). Fig.~\ref{fig:froc_illustration} illustrates how the detection hit is calculated in the FROC analysis. It is important to note that for objects with elongated shapes, the Intersection over Union (IoU) tends to vary. Consequently, any detection proposal with an IoU $\geq 0.2$ is considered a detection hit. Otherwise, it is marked as a false negative (a fracture on the ground truth that was not detected) or a false positive (a predicted fracture that does not exist in the ground truth). For each detection proposal, there is an associated predictive confidence (Conf). When calculating the FROC curve, for each FP level, a confidence threshold (Conf Thresh) needs to be determined. This threshold is used to distinguish whether a detection hit is a true positive or false negative for the current FP level.

The final detection metric used in the challenge ranking is the average of sensitivities at FP levels of 0.5, 1, 2, 4, and 8. In addition to the FROC analysis, we also calculate the maximum detection sensitivity (Max Sensitivity) and the average false positives per scan (Avg FP). These metrics represent the sensitivity and number of FP without considering the FP level. Although the clinical importance of segmentation is not as critical as detection, we also provide the IoU and Dice score for reference in segmentation predictions.

\subsubsection{Classification}
\begin{figure}
\centering
\includegraphics[width=\linewidth]{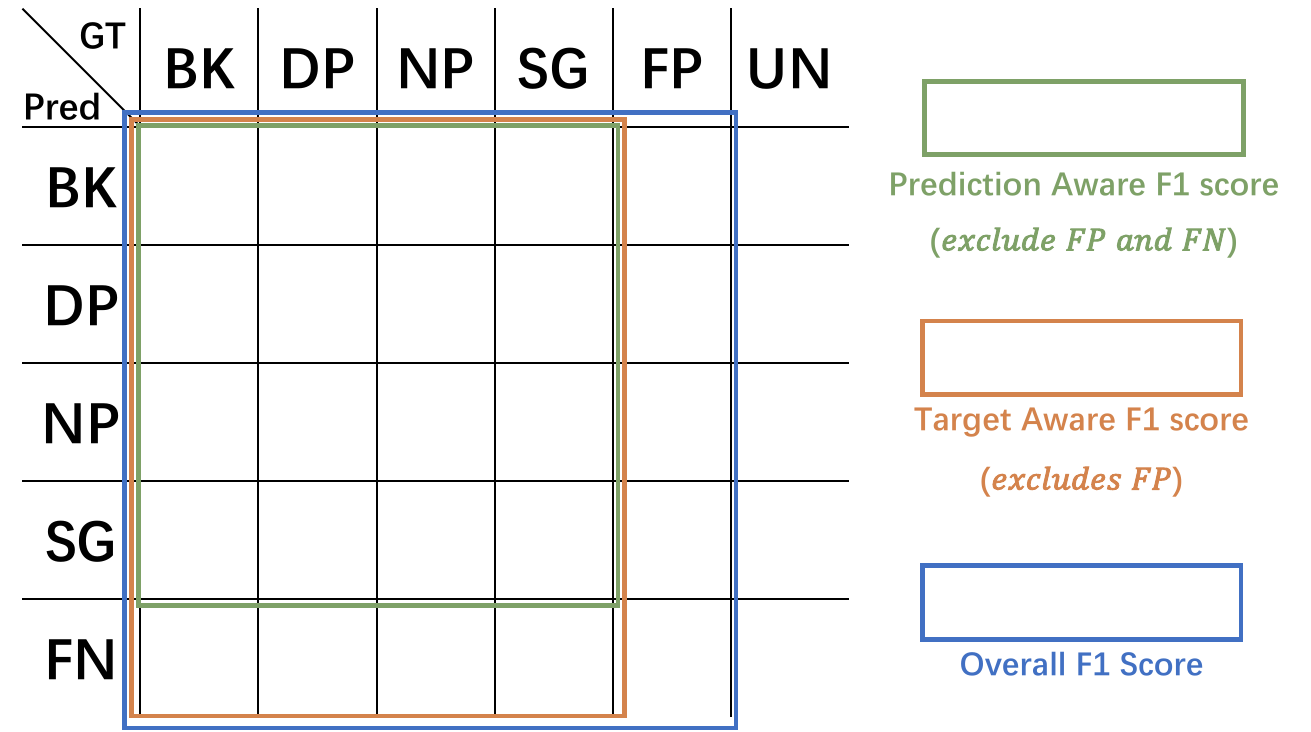}
\caption{\textbf{Illustration of classification confusion matrix and three F1 scores.} Overall F1 score (blue) evaluates the end-to-end classification performance (from detection to classification), target-aware F1 score (orange) evaluates performance on the classification annotations (excluding FP), and prediction-aware F1 score (green) evaluates performance on the classification predictions (excluding FP and FN). FN and FP refer to false negative and false positive predictions in detection.}
\label{fig:f1_illustration}
\end{figure}

In this task, participants are required to classify the detected rib fractures into 4 clinical classes (BK, ND, DP, and SG) based on the results of Task 1. During evaluation, the classification predictions are organized into a confusion matrix with 5 rows and 6 columns, as shown in Fig.~\ref{fig:f1_illustration}. The 5 rows represent the 5 categories of predictions: BK, ND, DP, SG, and FN (false negative of detection), while the 6 columns represent the 6 categories of ground truth: BK, ND, DP, SG, FP (false positive of detection), and UN (unclassified labels that are ignored during evaluation). The evaluation metric for classification is the macro-average F1 score, calculated using the confusion matrix. We compute 3 classification F1 metrics, each measuring different aspects of the classification system:

\begin{itemize}
\item \textbf{Overall F1} evaluates the overall classification performance, integrating it with the detection system. It is used as the final classification ranking metric.

\item \textbf{Target Aware F1} evaluates the classification performance based on the classification annotations, excluding all false positives.

\item \textbf{Prediction Aware F1} evaluates the classification performance based on the classification predictions, excluding all false positives and false negatives.
\end{itemize}


\subsection{Challenge Setup}

The RibFrac Challenge was held as part of MICCAI 2020, following the MICCAI Challenge guidelines and undergoing a rigorous review process that included two independent reviews and a meta review. The challenge design document is available on Zenodo~\cite{ribfracdocument}. The evaluation of participant performance was conducted using the established evaluation methods outlined in Section~\ref{sec:evaluation}. The ranking metric for the detection track was the average sensitivity in FROC analysis, while for the classification track, the overall F1 score was used to evaluate the end-to-end system. It is worth noting that although the performance of the classification track partially relied on the detection track, the two tasks were evaluated separately.

The challenge consisted of two phases. In Phase 1, participants were provided with training and validation images along with their corresponding labels. In Phase 2, test images were released, and participants were given a two-week period to submit their solutions. During Phase 2, only the ranking metrics were visible to the participants, which were used to determine the final challenge ranking.

Cash prizes and invitations to the challenge workshop were awarded to the top 3 teams in each track, resulting in a total of 5 teams (with one team overlapping). Additionally, 10 teams were invited to contribute to this paper, and ultimately 7 valid solutions were received.

Following the official MICCAI challenge period, the RibFrac Challenge continues to serve as an online benchmark. All the evaluation metrics, including those used for ranking as well as additional metrics, remain accessible to participants, facilitating continuous monitoring of new developments and enabling further improvements.


\section{Internal Rib Fracture Detection Methods}

\begin{figure*}
\centering
\includegraphics[width=0.8\linewidth]{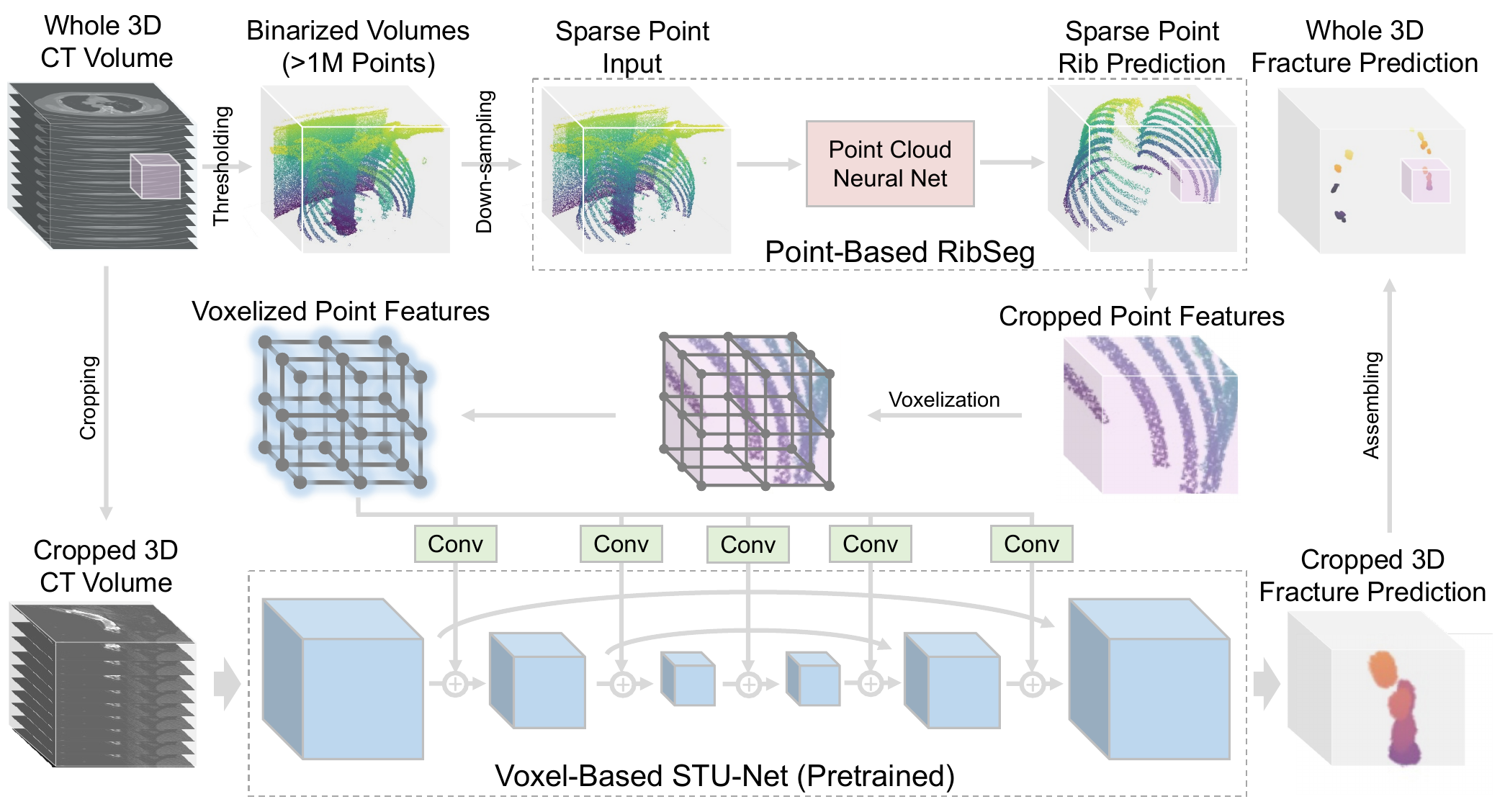}
\caption{\textbf{A schematic overview of the FracNet+ framework}, illustrating the dual-branch architecture for rib fracture segmentation. The top branch, the point-based rib segmentation network (RibSeg~\cite{Yang2021RibSegDA,Jin2022RibSegVA}), processes binarized and downsampled point clouds from the whole 3D CT volume to predict ribs. The bottom branch, a voxel-based rib fracture segmentation network (a pretrained STUNet~\cite{huang2023stu}), operates on cropped 3D CT volumes from rib areas to predict fractures. The voxelized point features from RibSeg are fused with the voxel-based fracture predictions to enhance the final fracture detection by integrating global context and local anatomical details.
}
\label{fig:internal_method}
\end{figure*}

In this section, we will introduce the development of our internal rib fracture detection method, FracNet+ as illustrated in Fig.~\ref{fig:internal_method}, which is an expansion of our previous FracNet model~\cite{Jin2020DeeplearningassistedDA}. We will begin by briefly reviewing the approach of FracNet, and then proceed to enhance its design by integrating two recent advancements in the field---the first is a point-based rib segmentation technique~\cite{Yang2021RibSegDA,Jin2022RibSegVA}, and the second involves the use of large-scale pre-trained networks for 3D medical imaging~\cite{huang2023stu}. Following this, we will delve into the model design details of the FracNet+ model.

\subsection{FracNet as an Internal Baseline}

In rib analysis, traditional methods primarily rely on the extraction of rib centerlines, followed by techniques such as rib unfolding~\cite{ringl2015ribs,Bier2015EnhancedRT} to convert original 3D images into 2D stacks for analysis. While this approach effectively utilizes anatomical knowledge of the ribs, it often encounters issues like false negatives due to the limitation of the analysis to two dimensions. Contrastingly, FracNet adopts a fully data-driven methodology. By moving away from the dependency on rib centerlines, FracNet overcomes the difficulties and inaccuracies involved in translating 3D structures into 2D representations, a common challenge in rib unfolding. This strategy aligns more closely with the complex, three-dimensional nature of rib fractures. Consequently, FracNet leverages the full spatial context provided by 3D CT scans for more accurate and reliable detection of fractures. This is especially pivotal in identifying subtle or complex fractures that might be missed or wrongly interpreted in 2D analyses. This approach is not only foundational to FracNet but has also become the basis for almost all detection solutions in the RibFrac Challenge, focusing on direct end-to-end learning and prediction of 3D instance segmentation of rib fractures.

The operational phases of FracNet include pre-processing, sliding-window prediction, and post-processing.

\begin{itemize}
    \item \textbf{Pre-processing.} This phase enhances detection efficiency by extracting bone areas using morphological operations like thresholding and filtering, preserving the original spacing of thin-section CT scans. Input voxel intensities are adjusted to a specific bone window (level=450, width=1100) and normalized to a range of $[-1,1]$.
    \item \textbf{Sliding-window prediction.} FracNet employs a specialized 3D UNet for rib fracture segmentation in a sliding-window manner. To accommodate the size of whole-volume scans, which may exceed standard GPU memory capacities, the scans are divided into $R\times R\times R$ patches with a $0.75R$-voxel stride, processed sequentially through the network. Here, $R$ denotes the size of the sliding window. The cumulative segmentation outcome is derived by assembling these individual patch predictions, maintaining the maximum values in overlapping areas.
    \item \textbf{Post-processing.} To reduce false positives, small predictions (under $200$ voxels) are excluded. Additionally, spine regions identified in the raw segmentation are removed. Detection proposals are generated by binarizing the post-processed segmentation at a low threshold ($0.1$) and computing connected components on this binary segmentation. Each connected component is considered a detection proposal, with its probability calculated from the average raw segmentation scores within the component.
\end{itemize}

\subsection{Extending FracNet to FracNet+}


\subsubsection{Integrated with Point-Based Rib Segmentation}

Although rib unfolding may not be the ideal utilization method, the ribs still have an anticipated role in fracture detection. Therefore, following the development of FracNet, we continue to evolve our methodologies by creating the RibSeg dataset and a point-based rib segmentation technique for rapid rib segmentation~\cite{Yang2021RibSegDA,Jin2022RibSegVA}. The basic concept is as follows:

We address the sparsity of ribs in 3D volumes, which comprise less than 0.5\% of the voxels, and the high Hounsfield Unit (HU) values characteristic of bones in CT scans. Our approach involves designing a point cloud-based model to segment ribs on binarized sparse voxels. Initially, we set a threshold of 200 HU to coarsely filter out non-bone voxels. These binarized volumes are then randomly downsampled and converted into point sets, simplifying computation before processing through the network. We utilize a point cloud neural net as the backbone and demonstrate robust performance in sparse 3D point cloud segmentation tasks. With RibSeg, we can swiftly segment ribs. For instance, by considering only a subset of points (30K), rib segmentation can be completed in under 1 second, in stark contrast to the over 70 seconds required by sliding-window based nnUNet~\cite{nnUNet}, while maintaining higher accuracy~\cite{Jin2022RibSegVA}.

A naive method to integrate RibSeg with FracNet is by replacing the bone extraction step in FracNet, which is based on morphological operations and can only roughly delineate bone areas, with a more precise rib extraction using RibSeg. This enables us to predict fractures using a sliding-window approach around the rib predictions, enhancing the accuracy and efficiency of the fracture detection process.


\begin{table*}
\scriptsize
\centering
\caption{\textbf{Solution summary in the rib fracture detection track}. Each of the solutions is summarized in the following aspects: the type of neural network (2.5D / 3D / Hybrid), pretraining, false positive reduction (FPR) / multi-stage (MS), backbone, input resolution, data augmentation methods, training loss, and special remarks. }
\label{tab:detection_solutions}
\begin{tabular}{m{0.06\linewidth}<{\centering}m{0.06\linewidth}<{\centering}m{0.06\linewidth}<{\centering}m{0.06\linewidth}<{\centering}m{0.1\linewidth}<{\centering}m{0.1\linewidth}<{\centering}m{0.13\linewidth}<{\centering}m{0.13\linewidth}<{\centering}m{0.09\linewidth}<{\centering}}
\toprule
Team & 2.5D / 3D & Pretraining & FPR / MS & Backbone & Input Resolution & Data Augmentation & Loss & Remarks \\
\midrule
lungseg2020 (DT1) & 2.5D  & ImageNet & Yes & ResNet50~\cite{7780459} & 800$\times$800$\times$15 & Random flip & Same as Mask R-CNN~\cite{2017Mask} (smooth L1, cross-entropy) & ImageNet pretraining \\
\midrule
DCC (DT2) & Hybrid & ImageNet & Yes & ResNet50~\cite{7780459}, ResNet101~\cite{7780459}, HRNet32~\cite{2020Deep} & 1,024$\times$1,024$\times$3, 128$\times$128$\times$32 & Random rotation and flip & Cross-entropy, smooth L1, Dice & ImageNet pretraining, Cascade  \\
\midrule
FakeDoctor (DT3) & 3D & No & Yes & 3D ResUNet~\cite{DIAKOGIANNIS202094} & 256$\times$256$\times$256 & Random rotation & Cross-entropy, Dice & Model ensemble \\
\midrule
UCIrvine (DT4) & 3D & No & No & 3D UNet~\cite{ronneberger2015u} & 64$\times$64$\times$64 & Random flip, rotation, scaling and transition & Cross-entropy, Dice & Model ensemble \\
\midrule
MIDILab (DT5) & 3D & No & No & 3D ResUNet~\cite{DIAKOGIANNIS202094} & 192$\times$192$\times$192 & None & Cross-entropy, Dice & Model ensemble \\
\midrule
CCCCS (DT6) & 2.5D & No & No & 2D UNet~\cite{ronneberger2015u} & 1,024$\times$1,024 & None & Dice & Preconditioning operations to extract the ribs \\
\bottomrule
\end{tabular}
\end{table*}

\subsubsection{Voxel-Point Fusion with Rib Segmentation}

Beyond the aforementioned naive integration, considering the guiding role of ribs in rib fracture segmentation, we propose a deep integration of features from the point-based rib segmentation network into the voxel-based fracture segmentation network. Specifically, our approach involves the following steps:

The FracNet+ network consists of two branches. The first branch is the point-based rib segmentation network, with an input-output mapping of \( N \times 3 \rightarrow N \times 1 \). This network processes a sparse point cloud sampled from the thresholded whole CT volume, outputting a score for each point indicating whether it is part of a rib. Here, \( N \) represents the number of input points. Based on the rib segmentation output, we sample a coordinate and expand a window of size \( R \times R \times R \) centered at it. Using this window, we crop the corresponding range of the original image to input into the second branch for fracture segmentation. Additionally, we crop the feature \( F_p \in \mathbb{R}^{M \times C_p} \) from the preceding layer of the output for subsequent feature integration, where $M$ and $C_p$ denote the number of the cropped point cloud and feature channel, respectively. 

The second branch is a voxel-based segmentation network with a UNet-like structure that inputs the cropped CT volume and outputs fracture segmentation scores for each voxel, \ie{}, \( 1 \times R \times R \times R \rightarrow 1 \times R \times R \times R \). At multiple stages of the voxel network, we integrate the feature \( F_p \) from the point branch. Specifically, for one of the voxel feature stages \( F_v \in \mathbb{R}^{C_v \times r \times r \times r} \), we first voxelize \( F_p \) into voxel features \( F'_p \in \mathbb{R}^{C_p \times r \times r \times r} \). In this step, features falling into the same voxel in the \( r \times r \times r \) grid are pooled together to form a single feature, which is a differentiable operation~\cite{liu2019point}. Afterwards, we use a \( 1 \times 1 \times 1 \) convolutional layer to transform \( F'_p \) into \( C_v \) channels, and then add it to \( F_v \). Mathematically,
$$
F'_p = \textit{Voxelization}(F_p) \in \mathbb{R}^{C_p \times r \times r \times r},
$$
$$
F'_v = F_v + \textit{Conv}(F'_p) \in \mathbb{R}^{C_v \times r \times r \times r}.
$$
The resulting feature \( F'_v \) combines the voxel branch features with the voxelized point branch features. Notably, although both \( F_v \) and \( F_p \) are derived from the cropped patch, \( F_p \) inherently possesses a global context, as its input is global. This means that the high-level information in the point cloud network has already aggregated features extending beyond the range of the cropped patch. Consequently, these enhanced features, particularly those emphasizing the rib area, can significantly improve the network's ability to segment fractures along the ribs. These fusion layers are applied across multiple stages of the voxel segmentation network. The integration of global and local features from both point and voxel branches in FracNet+ aims to capitalize on the strengths of both modalities, thereby enhancing the overall performance and accuracy of rib fracture segmentation.

\subsubsection{Large-Scale Pre-Trained Model}
A recent trend in medical imaging has been the development of large-scale pre-trained models, also referred to as medical large vision models~\cite{moor2023foundation,yang2023impact}. These models are trained on extensive datasets, enabling them to learn strong representations of medical images, which can significantly enhance their performance on specific tasks without extensive task-specific training. 

One of the key benefits of these large pre-trained models is their ability to act as a ``free lunch'' to boost the performance of applications in medical imaging. By leveraging the rich features learned from vast and varied data, they can provide a strong foundation for further fine-tuning on more specialized tasks, such as the segmentation of specific anatomical structures. In the context of our application, we have chosen STUNet~\cite{huang2023stu,wasserthal2023totalsegmentator}, a UNet-based model pre-trained on the TotalSegmentator dataset~\cite{wasserthal2023totalsegmentator}, pre-trained for the segmentation of 104 different organs from CT scans, notably including the ribs. It is worth mentioning that the rib labels in the TotalSegmentator~\cite{wasserthal2023totalsegmentator} were derived from models trained on our RibFrac dataset with RibSeg labels~\cite{Yang2021RibSegDA,Jin2022RibSegVA}. The choice of STUNet as the pre-trained model is strategic, as the extensive pre-training on large-scale CT scans with a broad range of anatomical structures provides a sophisticated understanding of the surrounding anatomy. To adapt STUNet for rib fracture detection, we employed a full model fine-tuning approach, as the annotations of rib fractures on the RibFrac dataset is sufficiently large.

\subsection{Model Details}

In this section, the method details will be outlined, encompassing the model architecture, training, and inference.

The FracNet+ architecture integrates two specialized branches: a point branch dedicated to rib segmentation and a voxel branch for rib fracture segmentation. The point branch is designed to work with any point cloud neural network as its backbone; in our implementation, we utilize PointNet++~\cite{qi2017pointnetpp} to process inputs of 30,000 points. For the voxel branch, we employ a pretrained STUNet Base~\cite{huang2023stu} ($R=128$) to maintain a balance between accuracy and computational efficiency. 

The two branches are jointly trained. It involves random point sampling for the point network and random cropping windows from rib for the voxel network. Both are trained using a combination of cross-entropy and Dice loss with rib and fracture segmentation as ground truth, respectively, without additional data augmentation beyond random sampling.

During inference, the FracNet+ first carries out the point branch computation to segment the ribs and caches the features from the last layer. Subsequent to this, center coordinates are sampled on the predicted rib segmentation to ensure that the sliding windows cover the entire rib area. The model then proceeds to the point-voxel fusion within these windows and predicts the rib fracture segmentation.







\section{Rib Fracture Detection Track}

\subsection{Challenge Solutions}


\begin{table*}[tb]
\centering
\caption{\textbf{A performance comparison of rib fracture detection task on the RibFrac test set}. We report the FROC metrics (\%), max sensitivity (\%) and avg false positives for rib fracture detection, IoU (\%) and Dice (\%) for rib fracture segmentation. The best metrics are highlighted in \textbf{bold}. The metric for ranking is highlighted in \colorbox{blue!25}{blue}.}
\label{tab:detection_performance}
\begin{tabular}{l|cccccc|cc|cc}
\toprule
\multicolumn{1}{l|}{\multirow{2}{*}{Method}} &
  \multicolumn{6}{c|}{Detection FROC (Sensitivities @FP levels)} &
  \multicolumn{2}{c|}{Detection Auxillary Metrics}  &
  \multicolumn{2}{c}{Segmentation} \\ 
\multicolumn{1}{l|}{} &
  \multicolumn{1}{c}{0.5 $\uparrow$} &
  \multicolumn{1}{c}{1 $\uparrow$} &
  \multicolumn{1}{c}{2 $\uparrow$} &
  \multicolumn{1}{c}{4 $\uparrow$} &
  \multicolumn{1}{c}{8 $\uparrow$} &
  \multicolumn{1}{c|}{\cellcolor{blue!25} Avg $\uparrow$} &
  \multicolumn{1}{c}{Max Sensitivity $\uparrow$} &
  \multicolumn{1}{c|}{Avg FP $\downarrow$} &
  \multicolumn{1}{c}{IoU $\uparrow$} &
  \multicolumn{1}{c}{Dice $\uparrow$} \\   
  \midrule
  \textit{Challenge Results} \\

lungseg2020 (DT1) & \bf 75.06 & \bf 79.74 & \bf 84.34 & \bf 87.06 & \bf 88.72 & \cellcolor{blue!25} \bf 82.98 & \bf 89.80 & 17.73 & 44.04 & 61.15 \\
DCC (DT2)     & 72.11 & 76.83 & 82.10 & 86.67 & 87.98 & \cellcolor{blue!25} 81.14 & 87.98 & 7.18  & 45.86 & 62.89 \\
FakeDoctor (DT3) & 56.74 & 75.01 & 80.69 & 84.36 & 84.81 & \cellcolor{blue!25} 76.32 & 84.81 & 5.84  & \bf 47.32 & \bf 64.24 \\
UCIrvine (DT4)  & 67.30 & 73.08 & 76.30 & 77.10 & 77.66 & \cellcolor{blue!25} 74.29 & 77.66 & 5.02  & 36.52 & 53.50 \\
MIDILab (DT5) & 60.09 & 65.01 & 68.48 & 69.84 & 69.84 & \cellcolor{blue!25} 66.65 & 69.84 & \bf 2.97  & 41.07 & 58.23 \\
CCCCS (DT6)& 53.29 & 61.19 & 67.57 & 72.25 & 76.00 & \cellcolor{blue!25} 66.06 & 80.61 & 29.10 & 23.15 & 37.60\\
\midrule
\textit{Post-Challenge Results}\\
A1 & 78.92 & 83.98 & 85.13 & 88.44 & 90.94 & 85.48 & 92.91 & 12.11 & 42.57 & 59.71 \\
A2 & 76.38 & 83.49 & 86.90 & 90.25 & 91.32 & 85.67 & 92.06 & 12.43 & 31.23 & 47.60 \\
A3 & 69.77 & 80.68 & 84.58 & 84.81 & 84.81 & 80.93 & 84.81 & 2.37 & 52.68 & 69.01 \\
\bottomrule
\end{tabular}
\end{table*}

The top solutions in the detection track are summarized in Tab.~\ref{tab:detection_solutions}. Among the six top solutions, three utilize 3D neural networks, two use 2.5D neural networks, and one proposes a hybrid pipeline with a 2D detection network and a 3D segmentation network. These solutions do not use any additional datasets. Most of the solutions do not employ any pretraining models, except for one solution that utilizes the ImageNet pretraining. The best-performing solution designs separate networks for false positive reduction, while another solution removes prediction masks with small areas to reduce false positives. The UNet~\cite{ronneberger2015u, DIAKOGIANNIS202094} architecture is widely used as the backbone in these solutions. For data augmentation, most solutions employ random flip or rotation, and some also apply random crop. Training losses include cross-entropy loss, Dice loss, smooth L1 loss, and soft IoU loss. Special techniques such as model ensemble are utilized to improve the detection performance. More method details for each team can be found here~\cite{ribfracsolutions}.

\subsection{Challenge Results}

The results of rib fracture detection for different solutions are presented in Tab.~\ref{tab:detection_performance}. Sensitivities at various false positive (FP) levels are reported for each solution, along with their average. The maximum sensitivity and average number of FPs are also provided as additional detection metrics, while Intersection over Union (IoU) and Dice scores serve as segmentation metrics. The average sensitivity at different FP levels is used as the final detection metric for ranking.

DT1 achieved the highest detection FROC score and maximum sensitivity.  Among the top solutions, DT5 had the fewest false positives but lower sensitivity. DT3 achieved the highest IoU and Dice scores, indicating superior segmentation performance. However, its FROC score was not as high, indicating possible calibration issues with the confidence of its detection proposals. This suggests that solely optimizing the segmentation network may not be the best strategy for the detection track, considering the low IoU threshold of 0.2 for detection proposals. It also highlights the effectiveness of false positive reduction.

\begin{figure}
\centering
\includegraphics[width=.9\linewidth]{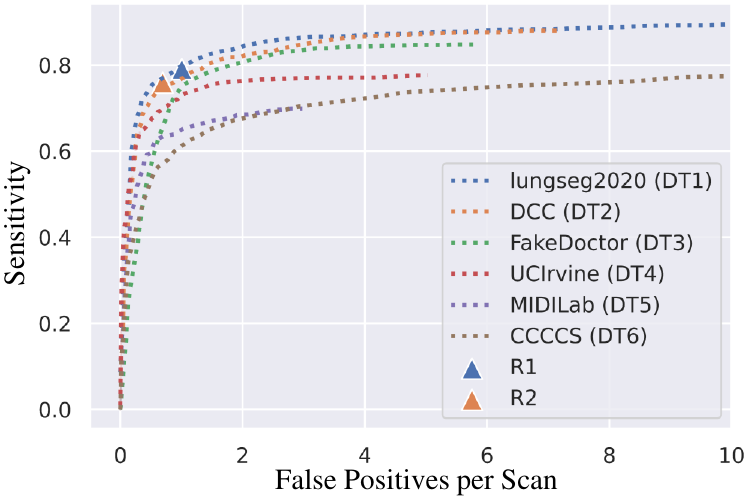}
\caption{\textbf{Performance comparison between top solutions and human experts.} FROC curves of each solution are displayed in dashed lines, and the performances of human experts are represented in triangles.}
\label{fig:froc_comparison}
\end{figure}

Fig.~\ref{fig:froc_comparison} illustrates a performance comparison between the top solutions and human experts. In an independent human-only observer study~\cite{Jin2020DeeplearningassistedDA}, a junior radiologist (R1) and a senior radiologist (R2), both highly experienced in chest interpretation, independently detected and segmented rib fractures in the RibFrac test set. The detection and segmentation metrics were computed based on the ground truth labels. The human experts achieved a detection false positive rate of 1.005 per scan with a sensitivity of 79.1\% (R1) and 0.690 per scan with a sensitivity of 75.9\% (R2). Despite the human experts having fewer false positives per scan, their detection sensitivities were lower compared to some of the top solutions. When considering the given false positive levels, the top two solutions perform comparably or even better than the human experts.

\subsection{Discussion}

\subsubsection{Analysis of Detection TP, FN and FP}


\begin{figure}
\centering
\includegraphics[width=\linewidth]{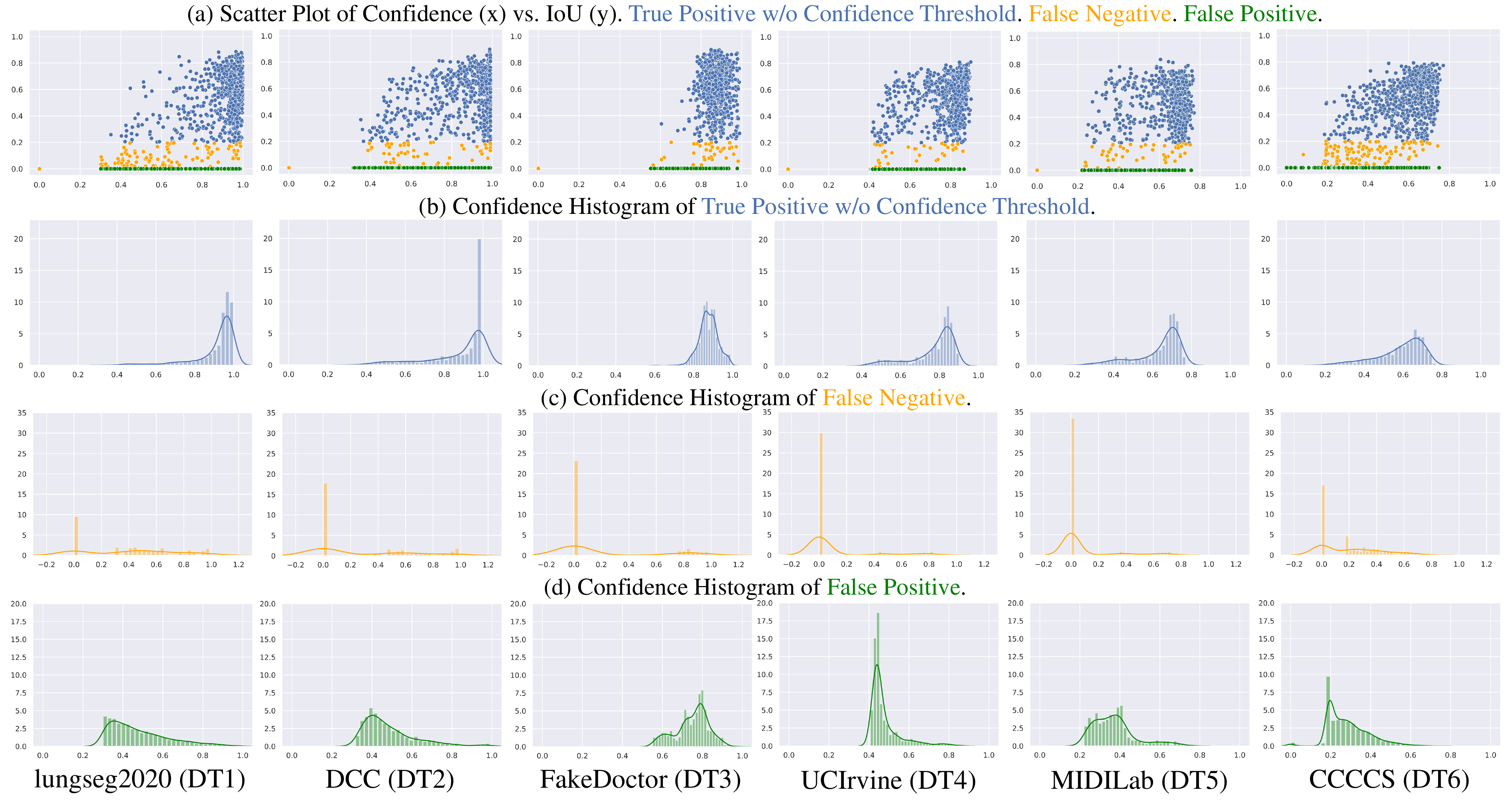}
\caption{\textbf{Statistics of detection confidence.} The scatter plots (a) illustrate the distribution of detection IoU (y-axis) of each prediction and its corresponding confidence (x-axis). The histograms further display the distribution of confidence score of (b) true positive predictions without confidence threshold, (c) false negative predictions, and (d) false positive predictions of each solution, respectively.}
\label{fig:detection_iou_conf}
\end{figure}

To analyze the relationship between the detection results (TP, FN, and FP), we present the scatter plot of confidence vs. IoU, the confidence histogram of TPs without a confidence threshold, and the histograms of FNs and FPs for each solution in Fig.~\ref{fig:detection_iou_conf}. Interestingly, most solutions demonstrate calibrated confidence values, where the confidence of TPs is generally high, while the confidence of both FNs and FPs is generally low. However, DT3 deviates from this trend, as it shows high confidence values for both TPs and FPs without significant discrimination. This precisely explains why this solution achieved high segmentation metrics but lower detection metrics. This analysis further emphasizes the importance of reducing false positives.

\subsubsection{Analysis and Visualization of Segmentation}


\begin{figure}
\centering
\includegraphics[width=\linewidth]{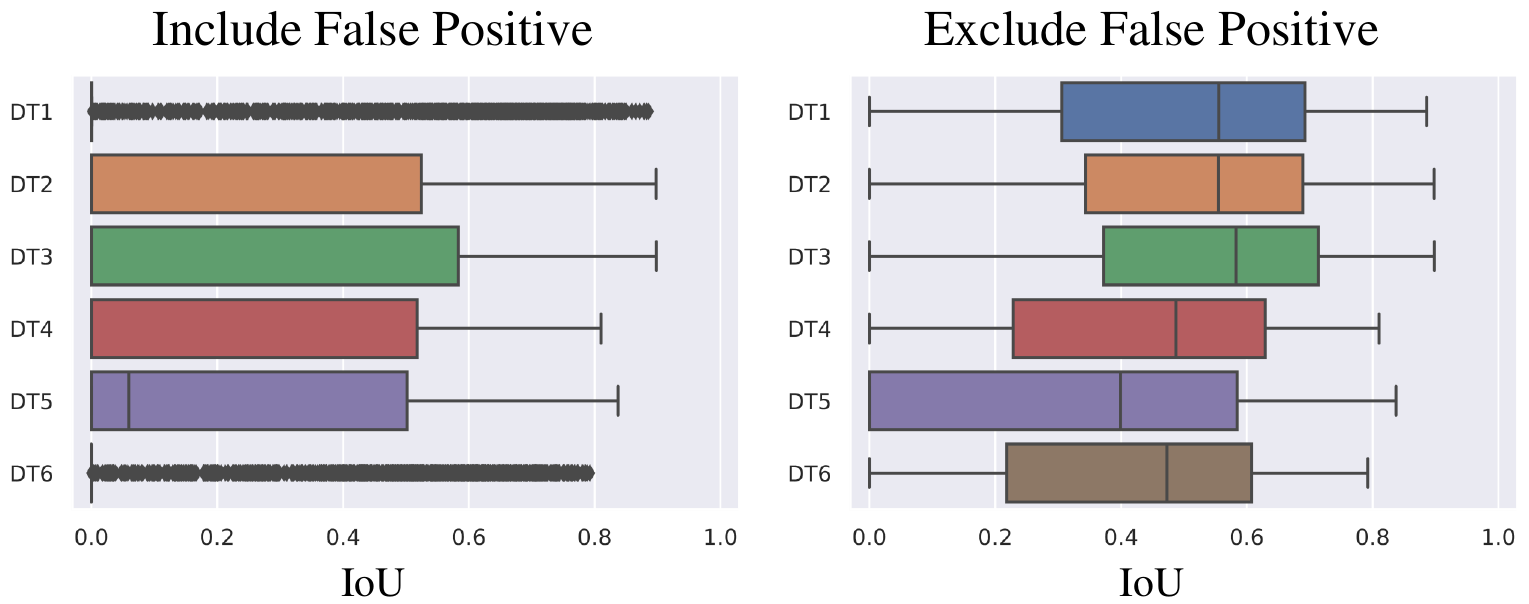}
\caption{\textbf{Statistics of segmentation performance.} The box plots display the distribution of IoU of segmentation including false positives (left) and excluding false positives (right).}
\label{fig:seg_iou}
\end{figure}

\begin{figure}
\centering
\includegraphics[width=\linewidth]{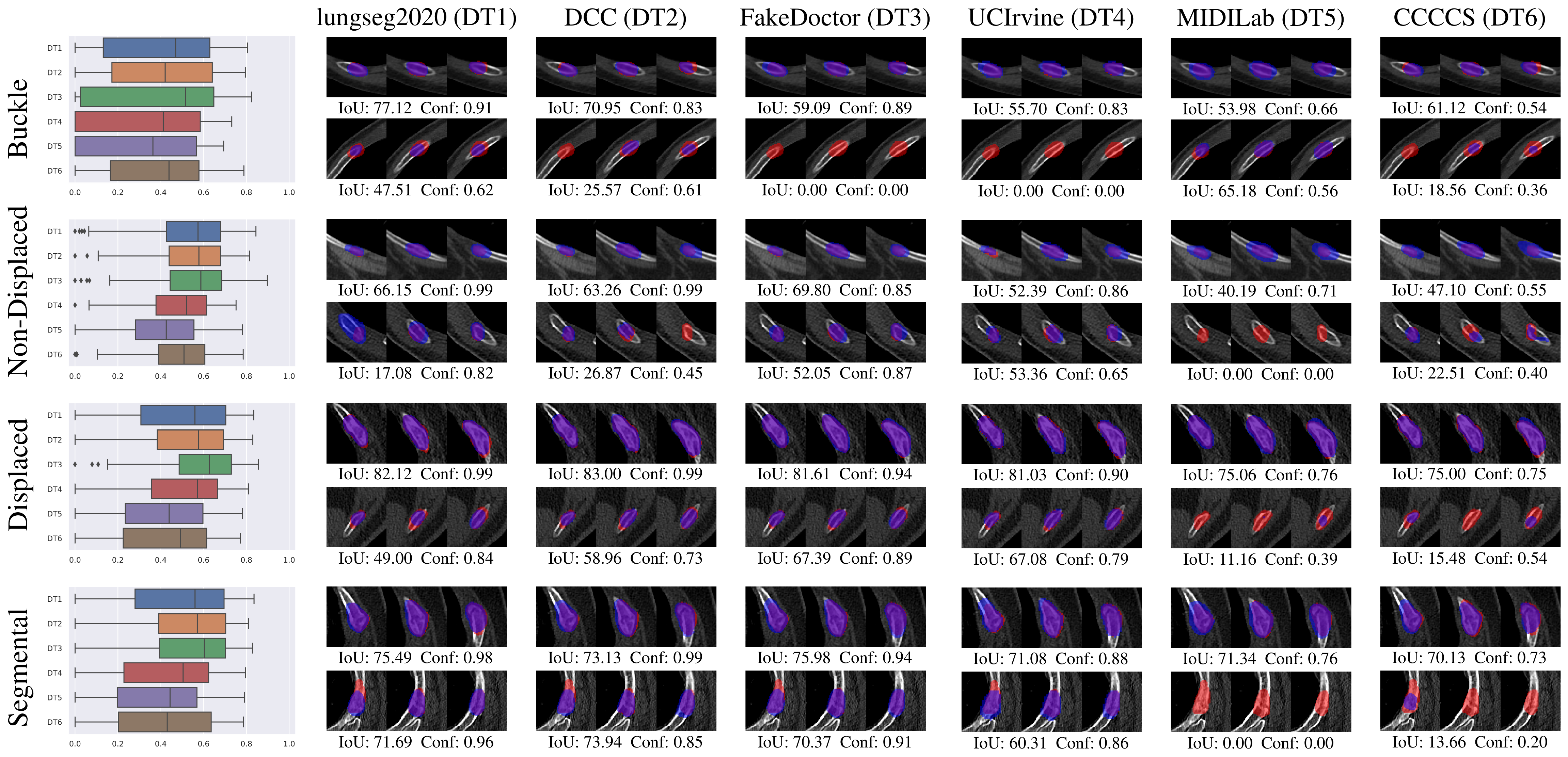}
\caption{\textbf{Statistics and visualization of segmentation performance for each category of rib fractures.} We provide a box plot showcasing the distribution of segmentation IoU values (excluding detection FPs) for each solution, along with visualizations of predictions for two representative cases. The ground truths are highlighted in red, while the corresponding predictions are depicted in blue. }
\label{fig:seg_visualization}
\end{figure}

We conducted an analysis of the segmentation performance using box plots, as shown in Fig.~\ref{fig:seg_iou}. The plots depict the distribution of IoU values for segmentation, with two variations—one including false positives (FPs) and the other excluding FPs. In the box plot that includes FPs, the segmentation predictions of DT1 and DT6 exhibit low IoU values, indicating a significant presence of false positives in their predictions. This observation is consistent with the average FP metric reported in Tab.~\ref{tab:detection_performance}. However, in the plot without FPs, the winning solution achieves high segmentation metrics. This suggests that the team prioritized maximizing sensitivity, which resulted in an increased number of FPs. Nevertheless, they were able to effectively suppress the confidence of these false positives through FPR, leading to a favorable FROC metric in the ranking, despite seemingly weaker segmentation metrics that were not for ranking.

Furthermore, in Fig.~\ref{fig:seg_visualization}, we present a box plot illustrating the distribution of segmentation IoU (disregarding detection FPs) for each category of rib fractures, along with visualizations of predictions for two cases in each category. For each visualization, we selected slices at fixed intervals around the center and annotated the ground truth segmentations in red, while the predictions are represented in blue. The analysis reveals varying levels of difficulty in segmenting different categories of rib fractures. For example, the overall segmentation IoU of buckle fractures is lower compared to that of non-displaced fractures. Contrary to the poor performance indicated in Tab.~\ref{tab:detection_performance}, the winning solution demonstrates satisfactory segmentation visualization results when false positives are disregarded.

\subsection{Post-Challenge Results}

As an ongoing online benchmark, the RibFrac Challenge has received numerous post-challenge submissions, some of which are on par with the top solutions during the official challenge period. 
As participant information is no longer required after the official challenge, we anonymized these submissions as A1, A2, and A3. Among them, the A1 method achieves better sensitivity levels at low FP level (FP=0.5), while A2 performs better at moderate FP level (FP=4). These advancements demonstrate that the RibFrac Challenge remains active and contributes to the development of methodologies.

On the other hand, we have observed limited progress in rib fracture classification. This is partly due to the inherent difficulty of rib fracture classification itself (as discussed in the analysis in Sec.~\ref{sec:classification-results}), and partly because the performance of the classification track in the RibFrac Challenge is heavily influenced by the detection performance, which can be overlooked by participants. We plan to address this issue to some extent in future versions of the RibFrac Challenge by modifying the evaluation methods, with the aim of promoting overall performance in the rib fracture classification task.

\subsection{Internal Experiments}


\begin{table*}[tb]
\centering
\caption{\textbf{Internal rib fracture detection methods on the RibFrac test set}. We report the FROC metrics (\%), max sensitivity (\%) and avg false positives for rib fracture detection, IoU (\%) and Dice (\%) for rib fracture segmentation. The best metrics are highlighted in \textbf{bold}. }
\label{tab:internal_method}
\begin{tabular}{l|cccccc|cc|cc}
\toprule
\multicolumn{1}{l|}{\multirow{2}{*}{Method}} &
  \multicolumn{6}{c|}{Detection FROC (Sensitivities @FP levels)} &
  \multicolumn{2}{c|}{Detection Auxillary Metrics}  &
  \multicolumn{2}{c}{Segmentation} \\ 
\multicolumn{1}{l|}{} &
  \multicolumn{1}{c}{0.5 $\uparrow$} &
  \multicolumn{1}{c}{1 $\uparrow$} &
  \multicolumn{1}{c}{2 $\uparrow$} &
  \multicolumn{1}{c}{4 $\uparrow$} &
  \multicolumn{1}{c}{8 $\uparrow$} &
  \multicolumn{1}{c|}{ Avg $\uparrow$} &
  \multicolumn{1}{c}{Max Sensitivity $\uparrow$} &
  \multicolumn{1}{c|}{Avg FP $\downarrow$} & 
  \multicolumn{1}{c}{IoU $\uparrow$} &
  \multicolumn{1}{c}{Dice $\uparrow$} \\   
  \midrule
  \textit{Internal Baselines} \\
  3D FCN~\cite{long2015fully} & 59.58 & 69.10 & 76.03 & 82.71 & 85.18 & 74.52 & 85.18 & 7.96 & 40.37 & 58.49 \\
  3D DeepLab~\cite{chen2018encoder} & 63.38 & 71.90 & 79.13 & 86.51 & 88.68 & 77.92 & 88.68 & 7.05 & 41.57 & 60.99 \\
  FracNet~\cite{Jin2020DeeplearningassistedDA} & 65.68 &	74.40 &	81.63 &	88.81 &	90.28 &	80.16 &	90.28 &	6.21 &	46.87 &	63.79  \\
\midrule
\textit{Extending FracNet to FracNet+} \\
FracNet w/ RibSeg (Naive)~\cite{Yang2021RibSegDA,Jin2022RibSegVA} & 69.90 & 75.96 & 82.52 & 89.50 & 90.58 & 81.69 & 89.65 & 4.91 & 48.69 & 65.77 \\
FracNet w/ RibSeg (Fusion)~\cite{Yang2021RibSegDA,Jin2022RibSegVA} & 73.56 & 78.22 & 83.86 & 89.92 & 90.77 & 83.27 & 91.29 & 4.34 & 49.59 & 66.81 \\

FracNet w/ STUNet (Random)~\cite{huang2023stu} & 66.20 & 75.04 & 82.40 & 89.46 & 90.94 & 80.81 & 90.83 & 5.89 & 47.15 & 64.10 \\
FracNet w/ STUNet (Pretrained)~\cite{huang2023stu} & 67.42 & 76.05 & 83.17 & 90.14 & 91.17 & 81.59 & 91.76 & 5.32 & 47.72 & 64.72 \\


FracNet+ & \bf 75.67 & \bf 79.22 & \bf 84.32 & \bf 90.82 & \bf 91.43 & \bf 84.29 & \bf 92.17 & \bf 4.22 & \bf 50.02 & \bf 67.22 \\
\midrule
\textit{Comparing with Human Experts} \\
R1 & - & - & - & - & - & - & 79.12 & 1.22 & 46.62 & 63.55 \\
R2 & - & - & - & - & - & - & 75.87 & 0.86 & 35.93 & 52.32 \\

\bottomrule
\end{tabular}
\end{table*}


In this section, we evaluated the performance of our internal methods, which include FracNet~\cite{Jin2020DeeplearningassistedDA}, FracNet+, and their variants for ablation study. This evaluation also encompassed models using only RibSeg, through either naive integration or point-voxel fusion, as well as models using only STUNet, whether with random initialization or pretrained on TotalSegmentator. It is crucial to highlight that all our internal methods are one-stage approaches that leverage a neural network for segmentation and determine the detection proposal confidence by calculating the average confidence across the segmented regions. Besides, the RibFrac Challenge test set includes 40 additional cases beyond the dataset utilized in \cite{Jin2020DeeplearningassistedDA}. This discrepancy accounts for the different metrics reported when compared to the original publication \cite{Jin2020DeeplearningassistedDA}.

The results in Tab.~\ref{tab:internal_method} indicate that while FracNet significantly outperforms 3D FCN~\cite{long2015fully} and 3D DeepLab~\cite{chen2018encoder}, FracNet+, an expansion of FracNet, demonstrates even greater effectiveness. This is evident whether through the integration of RibSeg or the use of the large-scale pretrained STUNet, both of which notably surpass the original FracNet.
Naive integration of rib outputs in pre- and post-processing, as seen in FracNet w/ RibSeg (Naive) vs. FracNet, also improves model performance, especially in reducing false positives, though it does not improve sensitivity. However, employing a point-voxel feature fusion to enhance rib features---FracNet w/ RibSeg (Fusion) vs. FracNet w/ RibSeg (Naive)---further boosts network performance, with a significant increase in sensitivity at low false positive levels and an overall rise in the highest sensitivity achieved.
On the other hand, substituting the voxel network with STUNet, even without using pretrained weights, shows a slight performance increase, as seen in FracNet w/ STUNet (Random) vs. FracNet. Introducing pretrained weights---FracNet w/ STUNet (Pretrained) vs. FracNet w/ STUNet (Random)---further improves overall performance. However, it is observed that merely changing the network structure or incorporating pretraining does not have as marked an effect as integrating RibSeg, as shown in FracNet w/ STUNet (Pretrained) vs. FracNet w/ RibSeg (Fusion), which highlights the usefulness of domain knowledge. Their effects seem to be complementary: while the former may further increase sensitivity, it does not significantly reduce false positives.
Finally, FracNet+, which is essentially FracNet w/ RibSeg (Fusion) w/ STUNet (Pretrained), combines the advantages of both modifications and achieves the best performance.

It is noteworthy that FracNet+ achieves higher sensitivity and finer segmentation compared to two human radiologists (R1, R2). Even though FracNet+ may yield more false positives than human experts, human radiologists can easily pinpoint lesions with fewer false positives, thereby enhancing the sensitivity of their diagnoses~\cite{Jin2020DeeplearningassistedDA}.

Additionally, though DT1 and DT2 surpass the performance of our earlier internal version FracNet, with the incorporation of recent advancements, FracNet+ demonstrates superior effectiveness. Even when compared to post-challenge results (A1 and A2), FracNet+, as a one-stage method, holds its ground in many metrics, particularly in segmentation accuracy and false positive rates. Considering the potential use of challenge tricks, such as multi-stage models (including false positive reduction) and ensemble techniques, there is room for further improvement in FracNet+. However, a direct comparison may not be entirely fair due to our later development timeline and complete access to the data as the authors. Therefore, we have refrained from employing too many challenge tricks and have chosen to present our internal results and challenge results in separate tables to maintain clarity and fairness. 

Nevertheless, our analysis indicates that while simply replacing the network may yield improvements, it does not achieve the integration with rib segmentation. This finding underscores the beneficial role of rib segmentation in enhancing fracture detection. 

\section{Rib Fracture Classification Track}

\subsection{Challenge Solutions}


\begin{table*}
\centering
\caption{\textbf{Solution summary in the rib fracture classification track}. Each of the solutions is summarized in the following aspects: the type of neural network (2.5D / 3D), pretraining, backbone, input resolution, data augmentation methods, training loss, and special remarks.}
\label{tab:classification_solutions}
\begin{tabular}{m{0.06\linewidth}
<{\centering}m{0.09\linewidth}
<{\centering}m{0.08\linewidth}
<{\centering}m{0.09\linewidth}<{\centering}m{0.11\linewidth}<{\centering}m{0.14\linewidth}<{\centering}m{0.10\linewidth}<{\centering}m{0.15\linewidth}<{\centering}}
\toprule
Team & 2.5D / 3D & Pretraining & Backbone & Input Resolution & Data Augmentation & Loss & Remarks \\
\midrule
UCIrvine (CT1) & 3D & No & NoduleNet~\cite{10.1007/978-3-030-32226-7_30} & 64$\times$64$\times$64 & Random flip, rotation, scaling and transition & Cross-entropy, Dice & Segmentation before classification \\
\midrule
DCC (CT2) & 3D & Med3D~\cite{chen2019med3d} & ResNet50~\cite{7780459} & 96$\times$96$\times$24 & Random rotation and flip & Cross-entropy  & Dilated convolutions~\cite{8354267}, auto context~\cite{5342426} \\
\midrule
DeepBlueAI (CT3) & 3D & No & Custom CNN & 64$\times$64$\times$64 & None & Cross-entropy, Dice & Segmentation before classification, SE~\cite{hu2018squeeze} \\
\midrule
CCCCS (CT4) & 3D & No & ResNet18~\cite{7780459} & 128$\times$128$\times$64 & None & Cross-entropy & Class-weighted loss \\
\bottomrule
\end{tabular}
\end{table*}


\begin{table*}[tb]
\scriptsize
\centering
\caption{\textbf{A performance comparison of rib fracture classification task on the RibFrac test set}. We report the prediction-aware F1 score, target-aware F1 score and overall F1 score for each category, as well as their macro-average. The best metrics are highlighted in \textbf{bold}. The metric for ranking is highlighted in \colorbox{blue!25}{blue}.}
\label{tab:classification_performance}
\begin{tabular}{l|ccccc|ccccc|ccccc}
\toprule
\multirow{2}{*}{Method} &
  \multicolumn{5}{c|}{Prediction-Aware F1 Score} &
  \multicolumn{5}{c|}{Target-Aware F1 Score} &
  \multicolumn{5}{c}{Overall F1 Score} \\
 &
  BK $\uparrow$&
  DP $\uparrow$&
  ND $\uparrow$&
  SG $\uparrow$&
  Avg $\uparrow$&
  BK $\uparrow$&
  DP $\uparrow$&
  ND $\uparrow$&
  SG $\uparrow$&
  Avg $\uparrow$&
  BK $\uparrow$&
  DP $\uparrow$&
  ND $\uparrow$&
  SG $\uparrow$&
  \cellcolor{blue!25} Avg $\uparrow$ \\
  \midrule
UCIrvine  & 0.3514 & 0.5503 & 0.5324 & 0.0833 & 0.3793 & 0.2114 & 0.5041 & 0.4744 & 0.0714 & 0.3153 & 0.2063 & 0.4844 & 0.4554 & 0.0690 & \cellcolor{blue!25} \bf 0.3038 \\
DCC       & 0.3478 & 0.4869 & 0.5249 & 0.0784 & 0.3595 & 0.2697 & 0.4693 & 0.4969 & 0.0741 & \bf 0.3275 & 0.1791 & 0.3790 & 0.3980 & 0.0678 & \cellcolor{blue!25} 0.2560 \\
DeepBlueAI & 0.2500 & 0.6839 & 0.6099 & 0.2500 & \bf 0.4485 & 0.0597 & 0.5550 & 0.4433 & 0.0976 & 0.2889 & 0.0377 & 0.4309 & 0.3857 & 0.0784 & \cellcolor{blue!25} 0.2332 \\
CCCCS   & 0.0000 & 0.4356 & 0.7424 & 0.0690 & 0.3117 & 0.0000 & 0.3929 & 0.7143 & 0.0625 & 0.2924 & 0.0000 & 0.2619 & 0.5782 & 0.0625 & \cellcolor{blue!25} 0.2257 \\
\bottomrule
\end{tabular}
\end{table*}

Tab.~\ref{tab:classification_solutions} provides a summary of the top solutions in the classification track. All of the top solutions employ 3D neural networks without pretraining, except for one team that utilizes a 3D ResNet-50~\cite{7780459} pretrained on Med3D~\cite{chen2019med3d}. The input resolutions vary from $64\times64\times64$ to $128\times128\times64$. While the top two solutions employ random rotation and flip for data augmentation, the remaining solutions do not utilize any data augmentation methods. Cross-entropy loss is used as the training loss for all teams, with two teams also employing Dice loss as an additional loss function. To enhance classification performance, some teams employ special techniques such as dilated convolution~\cite{8354267}, auto-context mechanism~\cite{5342426}, and class-weighted loss. More method details for each team can be found here~\cite{ribfracsolutions}.

\subsection{Challenge Results} \label{sec:classification-results}

The results of rib fracture classification are presented in Tab.~\ref{tab:classification_performance}. We provide the prediction-aware F1 score, target-aware F1 score, and overall F1 score for each rib fracture category, as well as their macro-average. It should be noted that the prediction-aware, target-aware, and overall F1 scores may not be consistent across the solutions, as these three classification metrics are biased towards the detection performance. However, since the labels of the test set are completely hidden, the classification track emphasizes the evaluation of end-to-end systems. The macro-average overall F1 score across the four categories is used for ranking.

Upon examining all the solutions, it is observed that the DP and ND fracture types are relatively easy to distinguish, while the BK type is slightly more challenging. However, the SG type shows very poor discriminative performance. Overall, although these solutions outperform random guessing, unfortunately, they are still far from being clinically applicable. In the following sections, we will analyze why the task of rib fracture classification is so challenging.

\subsection{Discussion: Why the Fracture Classification is Hard}

\subsubsection{Diagnostic Challenges and Class Imbalance}

The diagnostic approaches for these four types of fractures vary significantly. Non-displaced fractures (ND) and displaced fractures (DP) both involve complete fractures, with the distinction being the presence or absence of bone displacement or misalignment. As a result, these two types of fractures are relatively conspicuous and can be diagnosed based on their geometric positions. The results indeed indicate a higher diagnostic performance for ND and DP fractures. On the other hand, buckle fractures (BK) are incomplete fractures that typically manifest as bulges in the rib. This makes them easily missed in radiology examinations, leading to a higher difficulty in diagnosis and an increased number of false negatives in detection. The results demonstrate a significant improvement in the numerical value of the prediction-aware F1 score for BK fractures compared to the target-aware and overall F1 scores, as false negatives are excluded from the prediction-aware calculation. Segmental fractures (SG), in contrast, are defined as severe injuries characterized by at least two separate complete fractures in the same rib. Diagnosing SG fractures solely based on local information is not feasible. However, all the methods in the challenge utilized local patches, which resulted in poor discriminative performance for the SG.

Moreover, due to the RibFrac dataset being collected from real clinical scenarios, there is an imbalance in the distribution of fracture categories (as shown in Table~\ref{tab:ribfrac_dataset_split}). This further adds to the difficulty of machine learning tasks.

\subsubsection{Geometric Complexity}


\begin{figure}
\centering
\includegraphics[width=\linewidth]{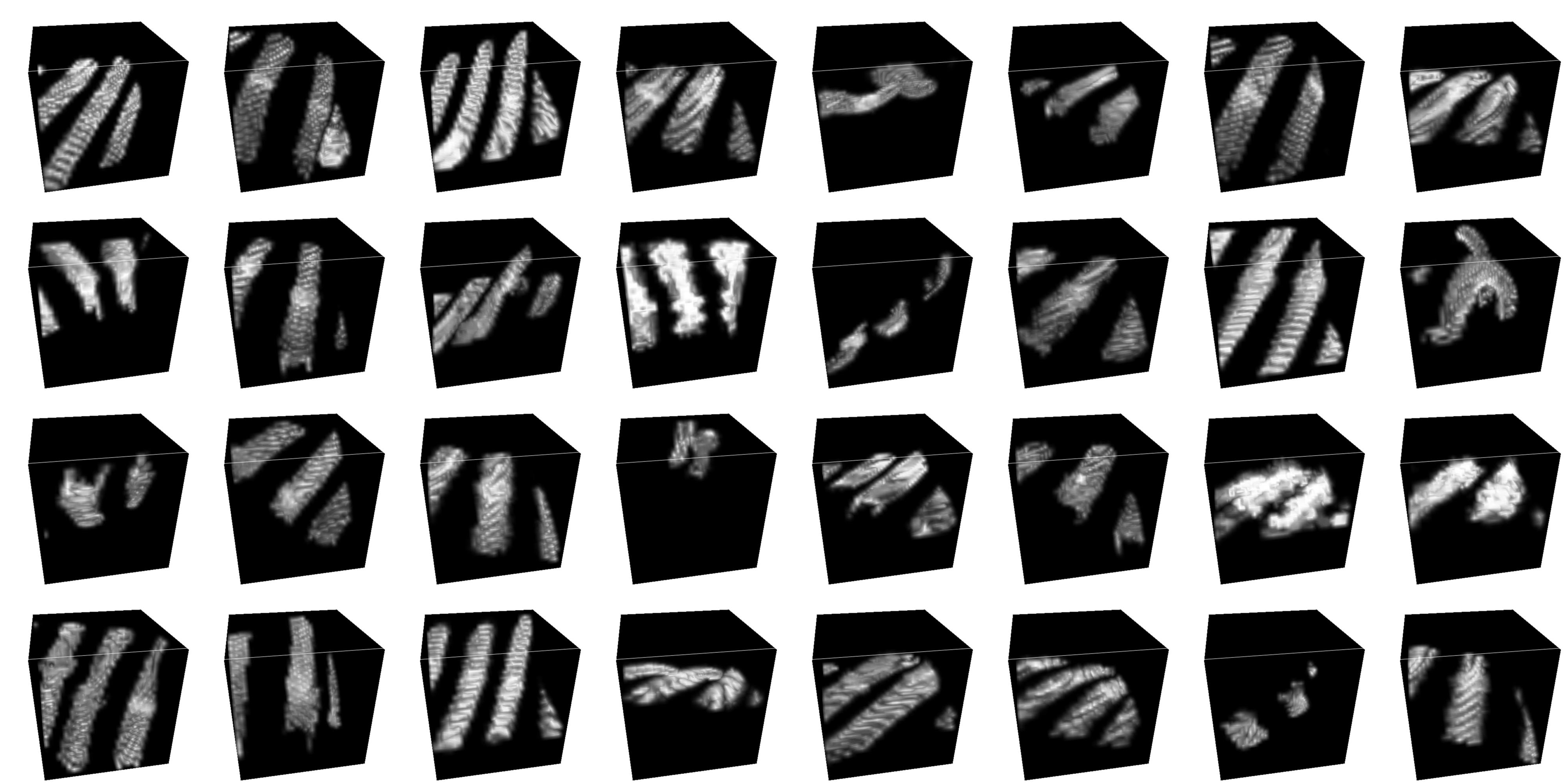}
\caption{\textbf{ Examples of rib fracture instances.} Each sample is center-cropped into a $64\times64\times64$ volume. }
\label{fig:geometric_difficulty}
\end{figure}

\begin{figure}
\centering
\includegraphics[width=\linewidth]{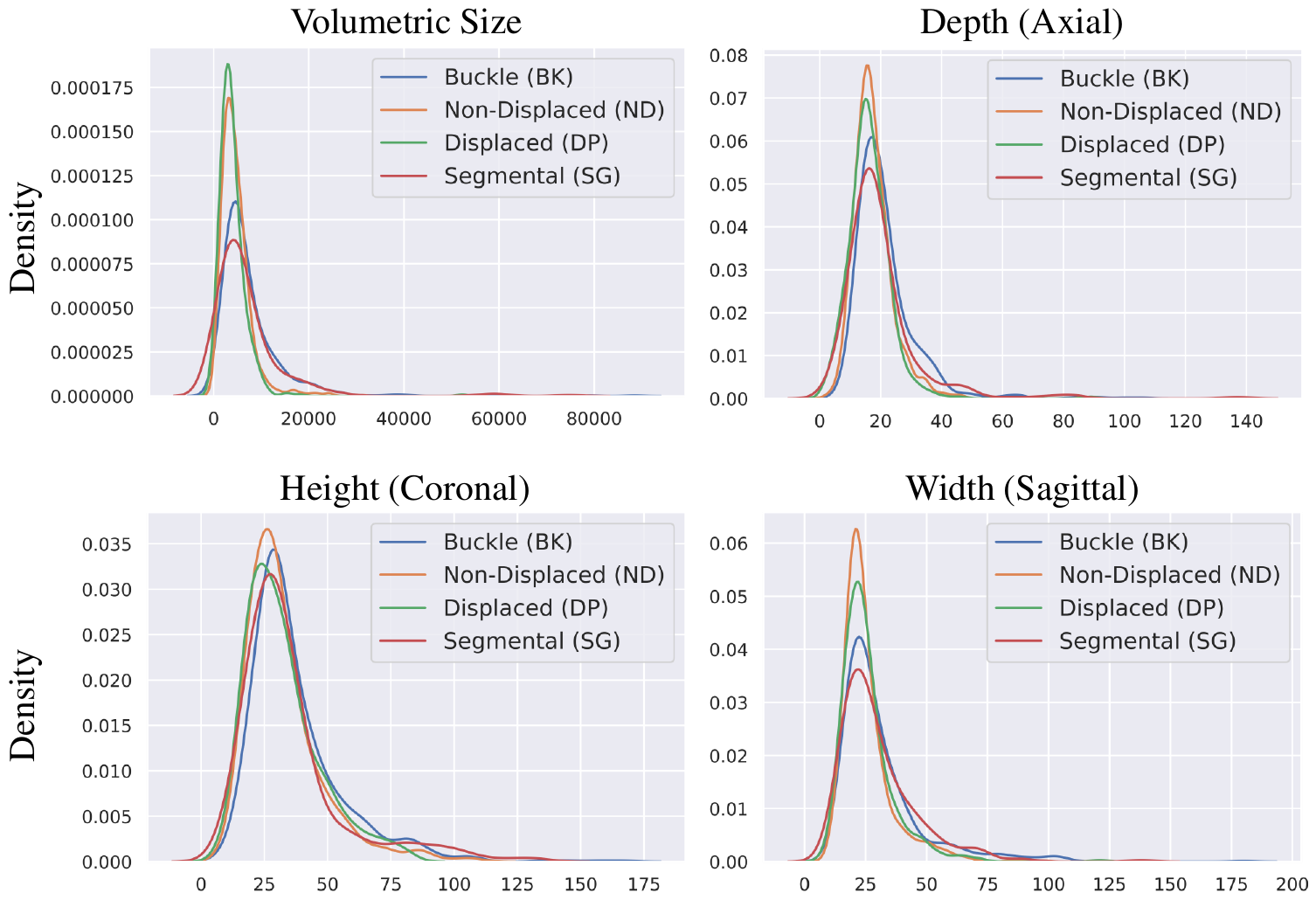}
\caption{\textbf{Statistics of sizes for each rib fracture category.} The histograms display the distribution of the volumetric size (top left), depth (top right), height (bottom left), and width (bottom right) of each individual rib fracture in each category. }
\label{fig:shape_variance}
\end{figure}

Rib fracture classification also faces the challenge of geometric complexity. Fig.~\ref{fig:geometric_difficulty} showcases several instances of rib fractures, where each sample is center-cropped into a $64\times64\times64$ volume. It can be observed that these 3D patches exhibit significant variations in orientation, shape, and appearance, posing a significant challenge for CNNs, which are sensitive to rotation. Despite the ideal assessment of rib fractures based on individual ribs, these 3D patches unavoidably contain multiple ribs, introducing additional background noise.

Furthermore, the sizes of the four fracture categories vary greatly. Fig.~\ref{fig:shape_variance} illustrates the distribution of volumetric size, depth, height, and width for each rib fracture category. These categories exhibit considerable size variation, and there is significant overlap between them, further complicating the classification task.


\section{Conclusion}

In conclusion, the RibFrac Challenge has served as a valuable benchmark and research resource for AI-assisted rib fracture detection and diagnosis. With its extensive dataset and evaluation platform, the challenge has addressed previous limitations and facilitated significant progress in rib fracture detection. Promising results have been achieved, showing that AI can enhance accuracy and reduce interpretation time. However, there is still room for improvement in rib fracture classification solutions. The challenge has paved the way for future research and integration of AI tools into radiology workflows. Overall, the RibFrac Challenge has played a crucial role in advancing the field and will continue to drive improvements in rib fracture analysis.

As an independent technical contribution, we develop a strong internal method FracNet+. We integrate several post-challenge advancements, including the development of a point-based rib segmentation technique and the emergence of large-scale pre-trained networks for 3D medical imaging, yielding competitive results in rib fracture detection and highlighting the importance of rib segmentation in this field.

Future research directions should concentrate on several areas. A primary goal is to develop unified approaches that seamlessly integrate rib labeling, anatomical centerline extraction, and fracture diagnosis. Moreover, enhancing the classification performance of different types of rib fractures is crucial. Despite advancements in fracture detection, accurately classifying various fracture types remains a complex challenge. Insights gained from this study of technical hurdles and existing methodologies will be invaluable for future endeavors in this aspect of rib fracture classification. Additionally, the development and validation of these models on multi-center data are of high importance. Addressing these future directions will not only expedite faster and more precise diagnoses but also enhance patient outcomes by enabling clinicians to make well-informed treatment decisions based on the severity and specific type of rib fractures. Continued collaboration, dataset expansion, and advancements in deep learning techniques will play vital roles in driving progress in rib fracture analysis and bringing us closer to clinically applicable solutions.



\appendix

\setcounter{table}{0}
\renewcommand{\thetable}{A\arabic{table}}

\setcounter{figure}{0}
\renewcommand{\thefigure}{A\arabic{figure}}

\subsection{Summary of Rib Fracture Detection Solutions}

\subsubsection{Team lungseg2020 (DT1)}
Mask-RCNN \cite{2017Mask} with FPN \cite{lin2017feature} and ResNet-50 \cite{7780459} backbone is extended from 2D to 2.5D for rib fracture detection and segmentation. Since it is a 2.5D method, the 3D CT scans are split into slices, resulting in 44,000 positive samples and 11,000 negative samples in the training set. Positive and negative samples are randomly selected with a 1:1 ratio during training. Horizontal flip is employed for training time augmentation, and the model is pretrained on ImageNet. Considering the relevance of adjacent slices in 3D CT scans, multiple adjacent slices are used as input to capture more contextual information, while the network outputs the prediction result of the middle slice. Experiments on the relationship between model performance and the number of adjacent slices input are conducted, and it is found that 15 is the optimal choice. During inference, each slice in a CT scan is processed independently. Predictions below a certain threshold are disregarded. 2D predictions are merged into 3D based on connectivity, and the 3D prediction score is calculated as the average of the 2D prediction scores.

\subsubsection{Team DCC (DT2)}
A cascaded rib fracture detection pipeline is proposed, consisting of a 2D slice-level detection network and a 3D patch-level segmentation network. The 2D detection network uses the Mask-RCNN \cite{2017Mask} model with ResNet \cite{7780459} and HRNet \cite{2020Deep} pretrained on ImageNet as the backbone. FPN \cite{lin2017feature} structure is added to enable multi-scale prediction. Random horizontal flip augmentation is applied to make the model learn invariant features. For the 3D segmentation network, UNet \cite{ronneberger2015u} is employed to capture stereoscopic features using an iterative training strategy. The predicted masks of Mask-RCNN are combined with the ground truth to obtain the training samples for UNet. The CT scans are cropped according to the center of each connected component before being fed into the network. Random rotation and flipping operations are performed for data augmentation. For model finetuning, predicted masks are integrated with ground truths to obtain new training samples.

\subsubsection{Team FakeDoctor (DT3)}
A two-stage rib fracture detection pipeline is proposed. It uses a ResUNet \cite{DIAKOGIANNIS202094} structure for rib fracture segmentation, followed by a classification model to predict the segmentation results. The down-sampling stage of the nnUNet \cite{nnUNet} model framework is replaced with ResUNet. Attention module and inflated module are applied to enlarge the subsampling receptive field. Combined binary cross-entropy loss and Dice loss are used as the loss function. In the post-processing stage of the segmentation model, prediction masks with smaller areas than the threshold are eliminated, while the remaining prediction masks are expanded to be slightly larger than the fracture area. In the classification model stage, the predicted masks are cropped and normalized through an adaptive pooling layer before being fed into a VGG \cite{vgg} network. The confidence level of each prediction mask is obtained after classification. Conventional data augmentation methods such as random rotation, translation, cutting, and blurring are employed.

\subsubsection{Team UCIrvine (DT4)}
A segmentation-based model with model ensembling techniques is used for rib fracture detection. Two 3D UNet \cite{ronneberger2015u} models are ensembled, one using binary cross-entropy loss and the other using Dice loss. The probability map is filtered with a threshold of 0.35, and each connected component is considered as an instance.

\subsubsection{Team MIDILab (DT5)}
An ensemble of 3D UNets \cite{ronneberger2015u} followed by morphological operations is applied for rib fracture detection. The model is based on a residual 3D UNet \cite{lee2017superhuman}, where all skip connections use element-wise addition instead of concatenation, and each module contains its own skip connection. The segmentation result is the voxel-wise mean of two residual 3D UNets, one trained on binary cross-entropy loss and the other on Dice loss \cite{crum2006generalized,sudre2017generalised}. For post-processing, each prediction is upsampled to the dimension of the original CT scans. The prediction is binarized with a threshold, and missing voxels in connected fracture regions are filled in with a morphological binary closing operation. Small predicted fracture instances smaller than 512 $mm^3$ are removed to avoid over-prediction. In the final binary prediction, instance labels are assigned by identifying and numbering the connected components. The post-processing hyperparameters that maximize the validation FROC are determined by a grid search and used in all future runs.

\subsubsection{Team CCCCS (DT6)}
A 2D residual UNet \cite{ronneberger2015u} is employed to segment rib fractures on 2D slices, followed by splicing the predictions into 3D as the post-processing operation. For data pre-processing, the ribs and their surrounding areas are extracted as input to address the problem of class label imbalance. The coronal slices are binarized with a threshold of 110 HU. A morphological dilation operation is applied to preserve soft tissues around the ribs. The binary images serve as the region-of-interest (ROI) masks, which are multiplied with the slices to obtain training data. Two slices of each annotation at the head and tail in the axial plane are disregarded to ensure that the rib fractures are clearly visible. The network architecture is a residual 2D UNet, with hybrid dilated convolution \cite{8354267} used to increase the receptive field. A two-stage training strategy is adopted to detect small rib fracture targets and reduce false-positive predictions simultaneously. Modified generalized Dice loss (GDL) \cite{10.1007/978-3-319-67558-9_28} is used to train a model with high recall, and then Tversky loss \cite{10.1007/978-3-319-67389-9_44} is employed for false-positive reduction. In the post-processing stage, a morphological dilation operation on spliced 2D slices in the axial plane is performed to eliminate crevices in predictions across slices.

\subsection{Summary of Rib Fracture Classification Solutions}

\subsubsection{Team UCIrvine (CT1)}
3D UNet~\cite{ronneberger2015u} with binary cross entropy is utilized for rib fracture segmentation. To limit the number of false positives, the probability map is filtered with a threshold of 0.95. For classification, the backbone of NoduleNet~\cite{10.1007/978-3-030-32226-7_30} is followed by three fully connected layers. A dropout layer is added between the last two fully connected layers to suppress overfitting. For data pre-processing, logarithm transformation is applied to the filtered and normalized CT scans, to balance out the effect of long tail data distribution. Each rib fracture in training and validation data is cropped into a 3D bounding cuboid with 4 extra pixels in each direction, with cuboid labeled -1 dropped. Random flipping, rotation, scaling and translation operations are conducted for training time data augmentation.

\subsubsection{Team DCC (CT2)}
3D ResNet-50~\cite{7780459} pretrained on Med3D~\cite{chen2019med3d} is used for rib fracture classification. Dilated convolutions~\cite{8354267} are applied in residual blocks to increase the receptive field while keeping the feature resolution. An auto-context mechanism~\cite{5342426} is adopted, where CT scans are integrated with masks generated by the segmentation model as the input of 3D ResNet-50 model. It demonstrated that the guidance of rib fracture mask could boost the classification performance. Random rotation and flipping operations are conducted to alleviate the over-fitting problem.

\subsubsection{Team DeepBlueAI (CT3)}
DeepMedic~\cite{Kamnitsas2017EfficientM3}, a dual pathway structure, is employed as the baseline of rib fracture detection. A 3D CNN with parallel convolution pathways extracts highly accurate and soft segmentation features, followed with a fully connected 3D conditional random fields (CRF) to generate segmentation labels. The dual pathways share the same structure but have different input resolutions, so that both local and contextual information are incorporated. Shortcuts connections are utilized in the structure, similar as those in ResNet~\cite{7780459}, to improve the training efficiency and boost performance. The segmentation features of the dual pathway 3D CNN are fed into fully connected layers, and then a 3D fully connected CRF with spatial regularization is used to generate smooth segmentation labels. Small connected components are removed for false-positive reduction.

\subsubsection{Team CCCCS (CT4)}
3D ResNet-18~\cite{7780459} with weighted cross-entropy loss is applied for rib fracture classification. The classification method is based on the segmented regions in rib fracture detection task, which is described in the detection track. The 3D variant version of ResNet-18 is used as the backbone. Due to the extreme imbalance between classes, weighted cross-entropy loss is adopted, with the normalized reciprocal of the number of samples in each category in the training set as the weight. 



\end{document}